\documentclass[structabstract]{aa}  
\usepackage[pdftex]{graphicx}
\usepackage{subfig,graphicx}
\usepackage{longtable,lscape,supertabular}
\usepackage{morefloats}
\usepackage{natbib}
\bibpunct{(}{)}{;}{a}{}{,} 
\begin{document}
   \title{
Searching Heavily Obscured Post-AGB Stars and Planetary Nebulae}

   \subtitle{I. IRAS Candidates with 2MASS PSC Counterparts}

\author{
G.\ Ramos-Larios\inst{1}, 
M.A.\ Guerrero\inst{1}, 
O.\ Su\'arez\inst{1,2}, 
L.F.\ Miranda\inst{1}, 
\and
J.F.\ G\'omez\inst{1} 
}

\institute{Instituto de Astrof\'{\i}sica de Andaluc\'{\i}a, CSIC. 
c$/$Camino Bajo de Hu\'etor 50, E-18008 Granada, Spain\\
\email{gerardo@iaa.es\thanks{Postdoctoral Research Fellow}, 
mar@iaa.es, 
lfm@iaa.es, 
jfg@iaa.es}
\and
UMR 6525 H.\ Fizeau, Universit\'e de Nice Sophia Antipolis, CNRS, OCA.  
Parc Valrose, F-06108 Nice Cedex 2, France\\
\email{olga.suarez@unice.fr}
             }

   \date{}

\authorrunning{Ramos-Larios et al.}
\titlerunning{
IRAS Obscured Post-AGB Star and PN Candidates with 2MASS Counterparts}

 
  \abstract
{
The transition from the Asymptotic Giant Branch (AGB) to the planetary 
nebula (PN) phase is critical in the shaping of PNe.  
It is suggested that the most asymmetric PNe are the descendant of 
massive AGB stars.  
Since these AGB stars are believed to evolve into heavily obscured 
post-AGB stars and PNe, the compilation of a sample of bona fide obscured 
post-AGB stars and PNe is important to help understand the formation 
of asymmetric PNe.  
}
{
We aim to identify and characterize in the IR a large number of 
post-AGB stars and PNe and to assess their degree of optical obscuration.  
The improved positions will enable future detailed studies to determine 
their true nature, whereas the optical and IR properties can be used 
to investigate their spectral behaviour.  
}
{
We have used 2MASS, \emph{Spitzer} GLIMPSE, \emph{MSX}, and \emph{IRAS} 
data in search of the near-IR counterparts of a sample of 165 presumably 
obscured \emph{IRAS} post-AGB and PN candidates, and DSS red images to 
identify the optical counterparts among the objects detected in the near-IR.  
The IR spectral energy distributions (SEDs) in the wavelength range 
from 1~$\mu$m to 100~$\mu$m of the sources with unambiguous near-IR 
counterparts have been analyzed using appropriate colour-colour 
diagrams.  
}
{
We have identified the near-IR counterparts of 119 sources out of the 
165 \emph{IRAS} post-AGB and PN candidates in our sample.  
The improved astrometric coordinates of these sources have allowed us to 
find optical counterparts for 59 of them, yielding a reduced sample of 
60 optically obscured post-AGB star and PN candidates.  
Among the 119 sources with near-IR counterparts, only 80 have unambiguous 
identifications in the 2MASS Point Source Catalogue.  
For these sources, we find that objects with and without optical counterpart, 
while having similar mid- and far-IR colours, are segregated in colour-colour 
diagrams that use the near-IR $J$ band to compute one of the colours.  
}
   {}

   \keywords{stars: AGB and post-AGB -- stars: evolution --
                planetary nebulae: general 
               }
   \maketitle
%

\section{Introduction}

Stars of low- and intermediate-mass ($0.8-1~M_\odot<M_i<8-10~M_\odot$) 
experience heavy mass loss episodes during the red giant phase that 
subsequently intensify at the tip of the asymptotic giant branch (AGB).  
With typical mass loss rates $10^{-4}-10^{-5}$ M$_\odot$~yr$^{-1}$, 
the AGB wind ejects most of the stellar envelope in a short time, 
and the star enters the post-AGB phase, from which it evolves into 
a planetary nebula (PN).  
The study of objects in the post-AGB phase, the short transition from 
the AGB to the PN phase, is essential to understand the transformation 
of a spherically symmetric AGB envelope to an aspherical PN.

Objects in the transition from the AGB to the PN phase are strong 
far-infrared emitters, making them easily detectable by the 
\emph{Infrared Astronomical Satellite} (\emph{IRAS}).  
Furthermore, they are located in the $[12]-[25]$ vs.\ $[25]-[60]$ 
\emph{IRAS} two-colour diagram in a well-defined region \citep{vdV88}. 
Therefore, the use of the \emph{IRAS} two-colour diagram provides 
an invaluable strategy that has resulted in a number of catalogues 
of young PNe and their immediate precursors, post-AGB stars 
\citep[e.g.][]{preitemartinez88,Petal88,vdV89,ratag90,hu93,pm07}. 
Recently, \citet{Setal06} presented a comprehensive optical spectroscopic 
atlas of post-AGB stars and young PNe selected from the \emph{IRAS} Point 
Source Catalogue.  
In this work, a number of post-AGB candidates were so heavily obscured 
that no optical counterparts were found, while some others were not 
observed because they showed signs of strong obscuration in the optical.

Heavily obscured post-AGB stars and PNe represent objects whose 
circumstellar envelopes have not expanded enough to become optically 
thin.
These objects may descend from the most massive AGB stars, as these 
are expected to be surrounded by very thick circumstellar envelopes during 
an important part or all the post-AGB evolution as a result of the large 
amounts of mass ejected in their envelopes \citep[up to $\sim$6~M$_{\sun}$ 
for a star with initial mass 7~M$_{\sun}$,][]{blocker95b}. 
Moreover, the rapid evolution of the most massive post-AGB stars 
do not provide sufficient time for the circumstellar envelope to 
expand and become optically thin. 
Evidences of this evolutionary path are provided by young PNe 
whose very thick envelopes result in extremely high internal 
extinctions that prevent their detection in the optical 
\citep[e.g., IRAS\,15103$-$5754,][]{vSP93,Setal06}.  
Obscured post-AGB stars might be key objects to understand the late 
evolution of the most massive PN progenitors.

With this work, we start a program devoted to search heavily obscured 
post-AGB stars and PNe and to characterize them in the IR.  
The observational material includes data available in the 2MASS 
database, the \emph{MSX} catalogue, and the \emph{Spitzer} Space 
Telescope GLIMPSE Galactic plane survey, as well as recently 
acquired more sensitive, higher resolution near-IR observations.
In this paper, the first in a series, we use 2MASS, \emph{MSX}, and 
\emph{Spitzer} GLIMPSE data to search for near-IR counterparts of 
presumably obscured \emph{IRAS} post-AGB star and PN candidates.  
The accurate location of the sources with 2MASS counterparts has allowed 
us to determine whether these sources are really obscured in the optical 
or not.  
The 2MASS, \emph{MSX}, and \emph{Spitzer} GLIMPSE data have been 
further used to characterize the near and mid-IR properties of 
the sources with 2MASS counterparts, allowing us to derive their 
spectral energy distributions (SED) in the IR domain.  
In a second paper, we will present new, high-quality near-IR {\it JHK} 
observations in order to search for the near-IR counterparts of the 
faintest objects, to find the near-IR counterparts of the objects not 
resolved by 2MASS, to describe the morphology of sources with extended 
emission, and to investigate the variability of the objects with 2MASS 
counterparts.

The sample and archival data used in this paper are presented in \S2 and 
\S3, respectively.  
The identification of the \emph{IRAS} sources with their 2MASS counterparts 
is described in \S4, and their spectral properties are discussed in \S5.  
Finally, a short summary is given in \S6.

\section{The Sample}

The sample of \emph{IRAS} post-AGB star and PN candidates investigated 
in this work is formed by the objects without optical counterpart in 
the sample of \citet{Setal06}, as well as by objects included in the 
original sample of these authors but not observed because they showed 
strong indications of heavy obscuration.  
These \emph{IRAS} sources fulfill the criteria used by \citet{Setal06} 
in order to select objects with a dust temperature in their envelopes 
of 80--200 K and an expanding radius of 0.01--0.1 pc.  
These selection criteria are:  
\begin{itemize}  
\item [(i)] 
The sources are well detected at 25 and 60 $\mu$m in the \emph{IRAS} 
Point Source Catalogue \citep[see][]{beichmancat88}. 
The flux quality for each band must be: 
  \begin{displaymath}
     {\rm FQUAL~(12~{\mu}m)\ge1;}  
  \end{displaymath}
  \begin{displaymath}
     {\rm FQUAL~(25~{\mu}m)=3;} 
  \end{displaymath}
  \begin{displaymath}
     {\rm FQUAL~(60~{\mu}m)=3.}
  \end{displaymath}

\item [(ii)] 
The ratios between the \emph{IRAS} photometric fluxes satisfy the 
following conditions:
  \begin{displaymath}
    \frac{F_{\nu}~(12~\rm\rm{\mu}m)}{F_{\nu}~(25~\rm{\mu}m)}\le0.50; 
  \end{displaymath}
  \begin{displaymath}
    \frac{F_{\nu}~(25~\rm{\mu}m)}{F_{\nu}~(60~\rm{\mu}m)}\ge0.35.
  \end{displaymath}

\item [(iii)] 
For sources well detected at 100 $\mu$m in the \emph{IRAS} Point Source 
Catalogue (FQUAL~(100~$\mu$m)$=$3), it is further imposed: 
  \begin{displaymath}
    \frac{F_{\nu}~(60~\rm{\mu}m)}{F_{\nu}~(100~\rm{\mu}m)}\ge0.60.
  \end{displaymath}

\item [(iv)]
The sources must show a low \emph{IRAS} variability index:
   \begin{displaymath}
     {\rm VAR}\le60\%
   \end{displaymath}

\end{itemize}

The region of the \emph{IRAS} two-colour diagram defined by the previous 
selection criteria has a small overlap with objects of different nature, 
mostly young stellar objects, OH stars, symbiotic stars, compact H~{\sc ii} 
regions, and Seyfert galaxies.  
In particular, the IR SEDs, and thus the IR colours, of the circumstellar 
material around young stellar objects \citep[e.g.][]{KH95} are very similar 
to these of post-AGB stars and young PNe, making it difficult to discern 
the nature of nebulae around young and evolved objects.  
On the contrary, Herbig Ae/Be stars, that exhibit distinct double-peak 
IR SEDs and are optically bright \citep[e.g.][]{Metal98,Metal01}, are 
not easily mistaken for post-AGB stars and young PNe.  
To minimize the number of young stellar objects in the sample, two 
additional criteria have been imposed to our sample:
\begin{itemize}
\item [(v)]
The sources are not classified in SIMBAD as young stars.  
\item [(vi)]
The sources are not located near the boundary of known star forming regions.  
\end{itemize}

\begin{figure}
\centering
    \includegraphics[width=0.95\columnwidth]{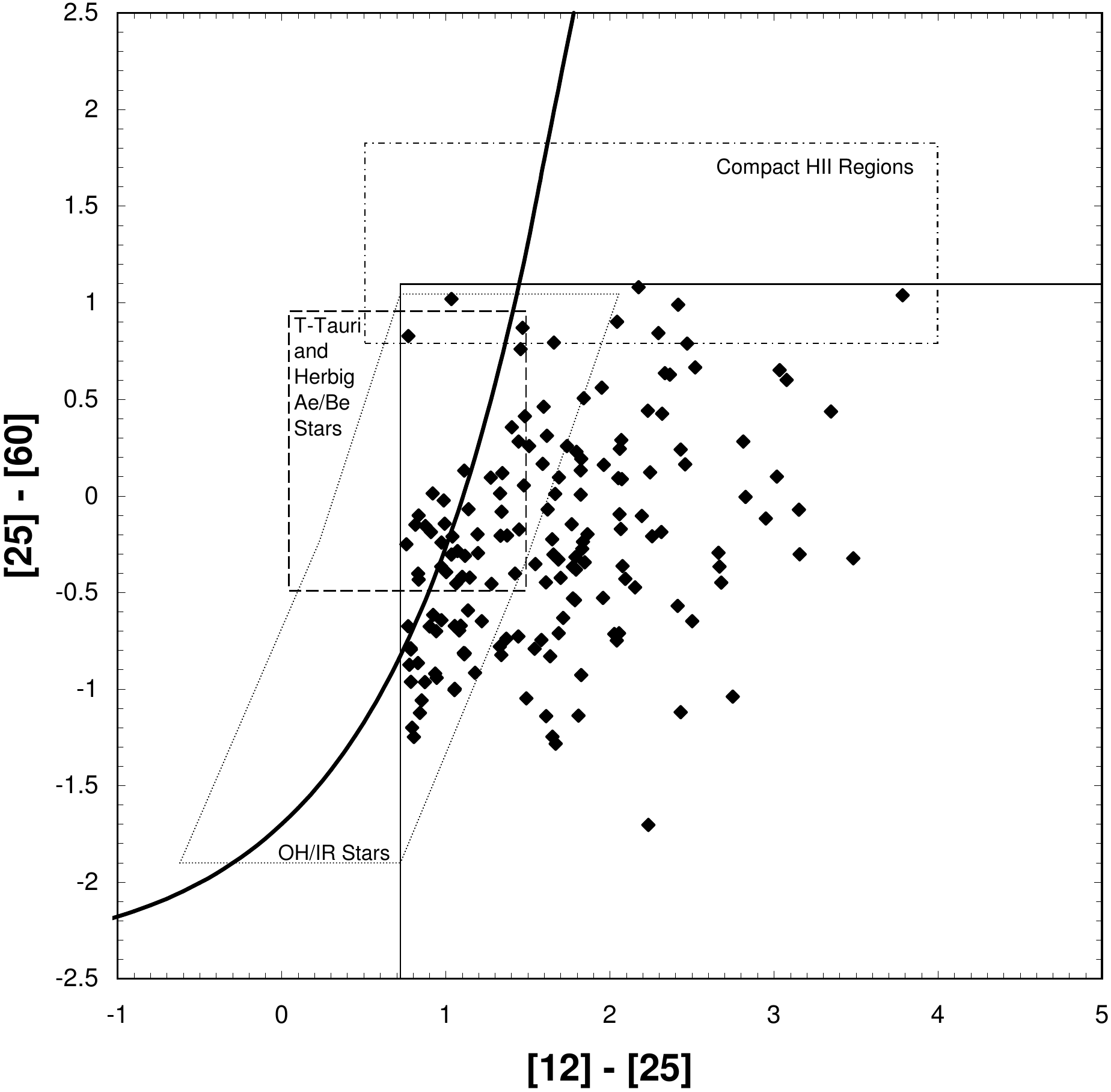}
\caption{
\emph{IRAS} colour-colour diagram showing the location of the \emph{IRAS} 
sources in our sample.  
The thin solid lines represent our selection criteria, while the 
solid thick line represents the expected colours for AGB stars 
\citep{B87}.  
The loci of OH/IR stars \citep{S89,tLetal91}, T-Tauri and Herbig Ae/Be 
stars \citep{HCH88}, and compact H~{\sc ii} regions \citep{AP87} that 
overlap with post-AGB stars and PNe are also overlaid on the diagram.  
\emph{IRAS} colours are computed as 
{[A]--[B]=2.5\,$\log(F_\nu({\rm B})/F_\nu({\rm A}))$}.  
}
\label{IRAS}
\end{figure}

With this criteria, the sample of \emph{IRAS} post-AGB star and young PN 
candidates being presumably obscured sources consists of the 165 objects 
listed in Table~1.  
This table includes the \emph{IRAS} names, coordinates, and error-ellipse 
(semi-major and semi-minor axes and position angle of the major axis) 
of these objects, together with their fluxes and flux quality factors
in the \emph{IRAS} 12~$\mu$m, 25~$\mu$m, 60~$\mu$m, and 100~$\mu$m
bands.  
The typical accuracy of the coordinates of the IRAS sources is
$\sim$25$\arcsec$, ranging from $\sim$10$\arcsec$ up to 45$\arcsec$
for the objects in Tab.~1.  
The uncertainty in the \emph{IRAS} coordinates is typically greater along 
the East-West direction.  
The location of these objects in the \emph{IRAS} two-colour diagram is
illustrated in Figure~\ref{IRAS}.

Some of the objects listed in Tab.~1 have been included in
previous studies.  
High-quality images or detailed studies are available for 
IRAS\,15452-5459 \citep{Setal07}, 
IRAS\,17150--3224 \citep[the Cotton Candy Nebula, ][]{KSH98,Hetal06}, 
IRAS\,13356--6249, IRAS\,13529--5934, IRAS\,15144--5812, IRAS\,17009--4154, 
and IRAS\,17088--4221 \citep{vSHW00}, and 
IRAS\,17516--2525 and IRAS\,18135--1456 \citep{Setal08}.  
Near-IR photometric observations of IRAS\,17404--2713, 
IRAS\,17487--1922, IRAS\,17516--2525, IRAS\,18071--1727, and 
IRAS\,19374+2359 \citep{vdV89}, 
and spectroscopic observations of IRAS\,17540--2753 and 
IRAS\,18135--1456 \citep{Oetal95} have also been reported.

\section{Archival Data}

In order to identify and characterize the near- and mid-IR counterparts 
of the \emph{IRAS} candidates to post-AGB stars and PNe in our sample, 
we have searched available 2MASS, \emph{MSX}, and \emph{Spitzer} GLIMPSE 
data products.  
The different properties of these catalogues are described below.

The 2 Microns All Sky Survey (2MASS) provides images of the complete 
sky 
in the $J$ (1.25 $\mu$m), $H$ (1.65 $\mu$m), and $K_{\rm S}$ (2.17 $\mu$m) 
bands \citep{Sketal06}. 
The image pixel scale is 1$\arcsec$ and the typical spatial resolution 
in the survey was $\sim$5$\arcsec$, as derived from the FWHM of the 
stars in the images. 
The 10 $\sigma$ detection limit for point-sources is $J<15.8$ mag, 
$H<15.1$ mag, and $K_{\rm S}<14.3$ mag. 

The Mid-course Space Experiment (\emph{MSX}) provides a detailed mid-IR 
map of a narrow strip, $|b| \leq 5^{\circ}$, of the whole Galactic plane in 
four different bands, namely band A (8.28 $\mu$m), C (12.13 $\mu$m), D 
(14.65 $\mu$m), and E (21.3 $\mu$m) .  
The positional uncertainty in the MSX Point Source Catalogue (MSX6C) 
is of the order of 2$\arcsec$, while the spatial resolution is quoted 
to be 18$\farcs$3 \citep{Pretal01}.  
The sensitivity levels for the bands A, C, D and E are 
160 mJy, 1200 mJy, 1000 mJy and 3000 mJy respectively.

The Galactic Legacy Infrared Midplane Survey Extraordinaire (GLIMPSE) 
project obtained \emph{Spitzer} Infrared Array Camera (IRAC) observations 
of the Galactic plane within the latitude and longitude ranges 
${|b| \leq 1^{\circ}}$, and ${|l| = 10^{\circ} - 65^{\circ}}$, 
respectively, 
in four bands with wavelengths (and bandwidths) 3.550 $\mu$m (0.75 
$\mu$m), 4.493 $\mu$m (1.9015 $\mu$m), 5.731 $\mu$m (1.425 $\mu$m), and 
7.872 $\mu$m (2.905 $\mu$m).  
These results were subsequently processed to yield the GLIMPSE 
Image Atlas 
with a pointing accuracy of $\sim$0$\farcs$3, and a spatial resolution 
that varies between 1$\farcs$7 and 2$\farcs$0 \citep{Fetal04}. 
The sensitivity (saturation levels) for the the different bands are 
0.5 mJy (439 mJy) in the 3.6 $\mu$m band, 
0.5 mJy (450 mJy) in the 4.5 $\mu$m band, 
2 mJy (2930 mJy) in the 5.8 $\mu$m band, and 
5 mJy (1590 mJy) in the 8.0 $\mu$m band.

\section{Source Identification}

The accuracy of the coordinates of an \emph{IRAS} source may not 
be good enough to allow a straightforward identification of its 
counterpart in the 2MASS, \emph{MSX} or GLIMPSE catalogues.  
This is especially the case of \emph{IRAS} sources 
in crowded fields located along the Galactic plane or near 
the Galactic Center, for which the error-ellipse typically 
includes several near-IR sources. 
In order to define a strategy to search for the near- and mid-IR 
counterparts of the \emph{IRAS} sources in Tab.~1, a comparison 
between the three databases used in this work can be useful.

The 2MASS images provide a complete coverage of the sky with a 
spatial resolution ($\sim$5$\arcsec$) better than the positional 
uncertainty of the \emph{IRAS} sources in Tab.~1.  
However, the sensitivity of the 2MASS images to obscured sources is 
limited because of the large extinction and intrinsic red colours of 
these sources.  
Accordingly, the identification of the 2MASS counterparts of the \emph{IRAS} 
sources in our sample is not obvious in general.  
This is not the case of the GLIMPSE Image Atlas that has a high 
sensitivity to mid-IR sources and provides better spatial resolution 
($\leq$2$\farcs$0) than the \emph{MSX} and 2MASS images.  
This database thus seems especially suited to determine with 
a great accuracy the position of the \emph{IRAS} sources in 
Tab.~1.  
Unfortunately, its limited spatial coverage results in a 
reduced number of sources with available GLIMPSE data. 
On the other hand, the \emph{MSX} data have an extremelly good sensitivity 
to obscured sources and its coverage is sufficiently good to include a 
significant fraction of the sources in our sample.  
The \emph{MSX} images, however, have a poor spatial resolution 
($\sim$18$\arcsec$) that can make difficult to pinpoint the 
near-IR counterpart.

\begin{figure*}[!t]
\centering
\includegraphics[width=0.8\columnwidth]{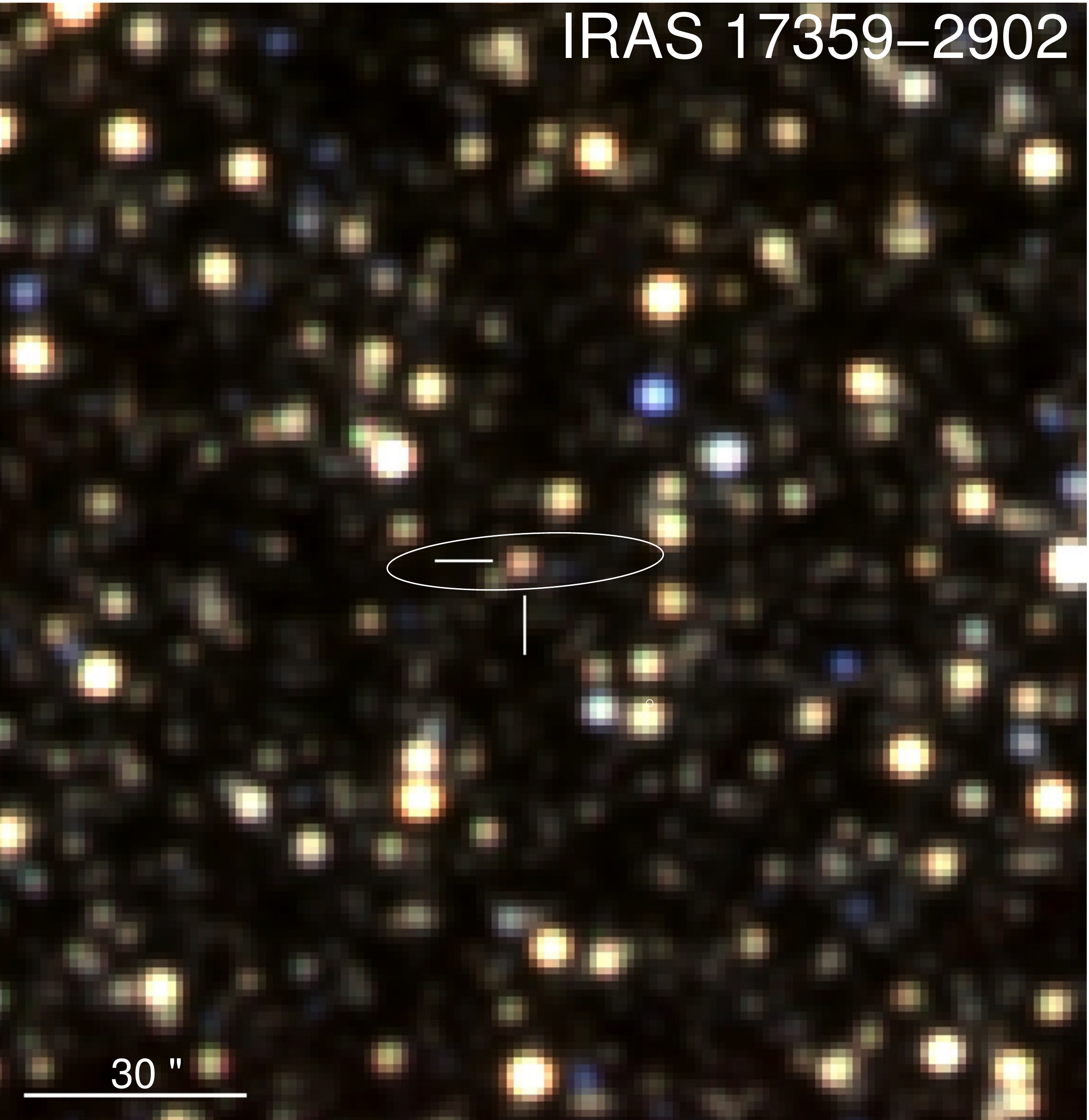}
\hspace*{\columnsep}%
\includegraphics[width=0.8\columnwidth]{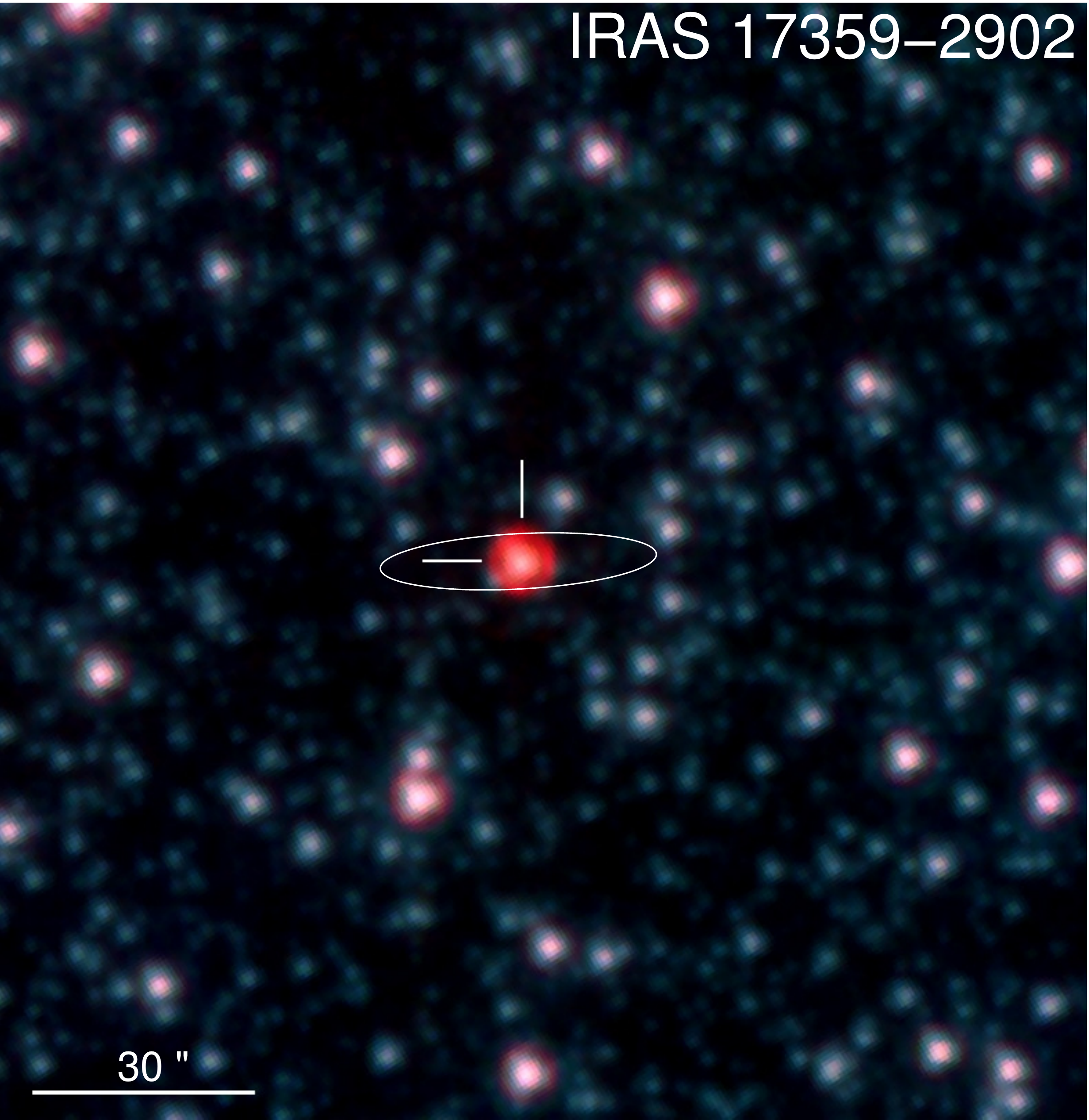}
\vskip .2in
\includegraphics[width=0.8\columnwidth]{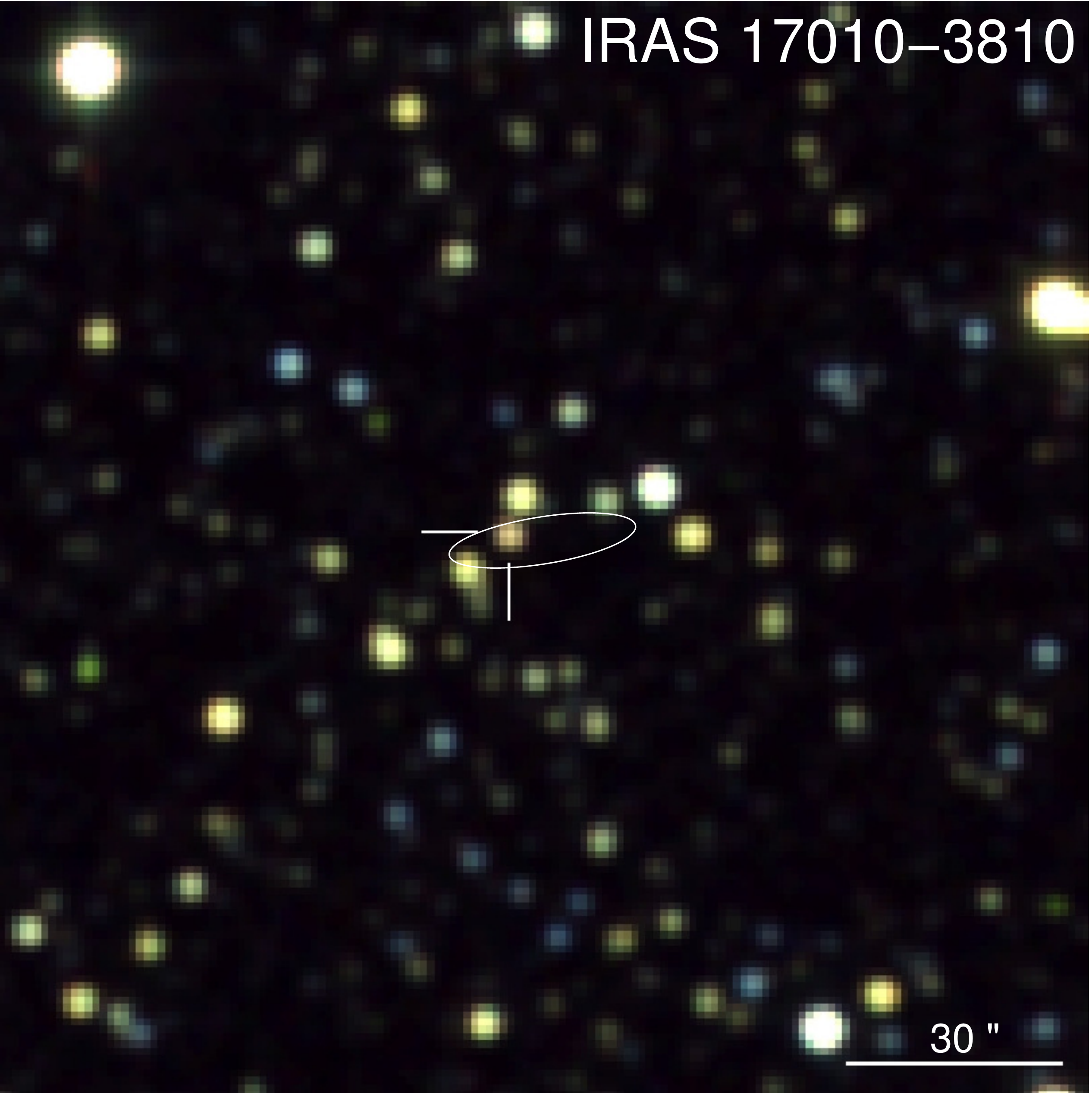}
\hspace*{\columnsep}%
\includegraphics[width=0.8\columnwidth]{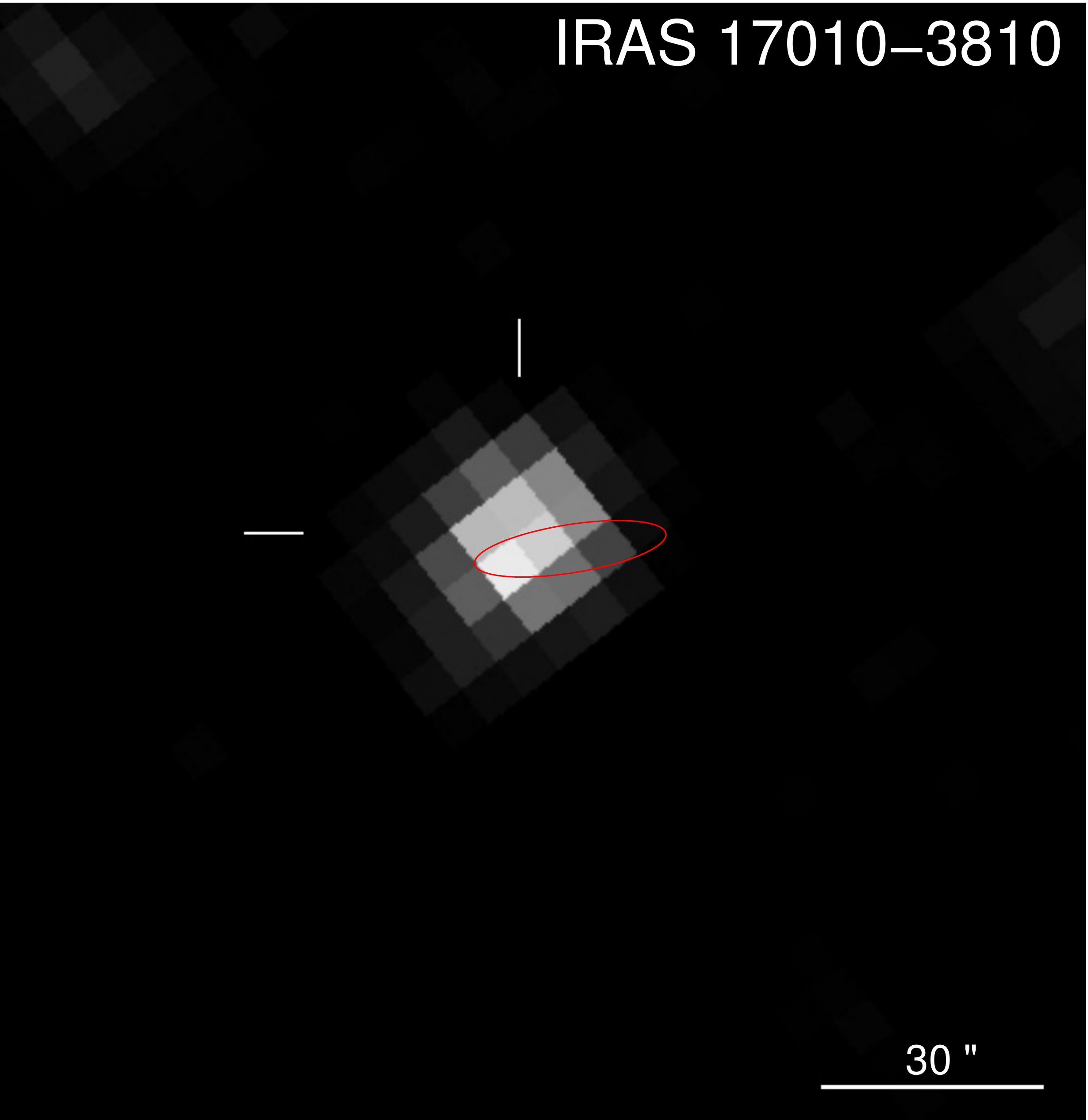}
\caption{
Representative examples of objects whose near-IR counterparts have been 
identified using \emph{Spitzer} GLIMPSE or \emph{MSX} data.  
({\it top}) 
2MASS $J$ (blue), $H$ (green), and $K_{\rm S}$ (red) composite picture 
({\it left}) and \emph{Spitzer} GLIMPSE 3.6 $\mu$m (blue), 4.5 $\mu$m 
(green), and 8.0 $\mu$m (red) composite picture ({\it right}) of 
IRAS\,17359$-$2902.  ({\it bottom}) 
2MASS $J$ (blue), $H$ (green), and $K_{\rm S}$ (red) composite 
picture ({\it left}) and \emph{MSX} image in the 8.28 $\mu$m A 
band ({\it right}) of IRAS\,17010$-$3810.  
The locations of the near-IR counterparts and the \emph{IRAS} ellipse-error 
are overlaid.  
North is up, east to the left.  
}
\label{ex}
\end{figure*}

Given the different properties of the 2MASS, \emph{MSX}, and GLIMPSE 
Image Atlas, we have proceed as follows in order to identify the near-IR 
counterparts of the \emph{IRAS} sources.  
When available, we have first examined the GLIMPSE images to pinpoint 
the accurate location of the \emph{IRAS} sources, thus allowing us to 
find its near-IR counterpart in 2MASS images.  
This procedure is illustrated in Figure~\ref{ex} for IRAS\,17359$-$2902; 
the bright \emph{Spitzer} mid-IR source inside its \emph{IRAS} error-ellipse 
allows the unambiguous identification of its near-IR counterpart among the 
near-IR sources within this ellipse in the 2MASS image.  
The use of the GLIMPSE images resulted in the identification of the 
near-IR counterparts of 22 sources in Tab.~1, although it must be 
noticed that some heavily obscured \emph{Spitzer} sources were not 
detected in the 2MASS images and are not listed in Tab.~1.  
We then examined the \emph{MSX} data of the sources with no GLIMPSE 
data available and were able to find the near-IR counterparts for 52 
additional sources in Tab.~1.  
A typical example of the identification of the near-IR counterpart of an 
\emph{IRAS} source using \emph{MSX} data is shown in Figure~\ref{ex} for 
IRAS\,17010$-$3810; the position of the \emph{MSX} source MSX6C G347.4963$+$01.8505, coincident within 1\arcsec\ with the position of the source 2MASS 17042731$-$3814417 
(marked in the 2MASS image), has allowed us to determine that this is the 
near-IR counterpart of IRAS\,17010$-$3810.  
Finally, we have examined the $JHK$ 2MASS images of the \emph{IRAS} sources 
with no GLIMPSE and \emph{MSX} data available, and selected the near-IR 
counterparts that appear as isolated sources within the error-ellipse of 
the \emph{IRAS} source and whose near-IR colours are typical of post-AGB 
stars or PNe, i.e., $(J-H)\sim1.0$~ and $(H-K)\sim0.4$ \citep{GLetal97}.  
This has added an additional number of 43 \emph{IRAS} sources 
with 2MASS counterparts.  
The result is a total of 119 sources with near-IR emission out of 
the sample of 165 sources listed in Tab.~1.

Among these 119 \emph{IRAS} sources detected in the near-IR, there are 
80 sources with a reliable counterpart in the 2MASS Point Source 
Catalogue (PSC).  
The remaning 39 \emph{IRAS} sources in Tab.~1 have near-IR emission 
detected in 2MASS, but the assignation of a counterpart in the 2MASS PSC 
is dubious due either to their extended appearance in the 2MASS images, 
to the possible contamination by an adjacent source, or to their extreme 
faintness, at the detection limit of 2MASS.

The sample of the 80 \emph{IRAS} post-AGB and PN candidates in Tab.~1 with 
a counterpart in the 2MASS PSC is listed in Table~2 with their coordinates 
and near-IR magnitudes, and their identification charts are shown in 
Figure~\ref{2MASSrgb1}.  
The angular separation with the position of the \emph{IRAS} 
source is also listed in this table.  
The distance between the \emph{IRAS} and 2MASS coordinates is typically below 
5$\arcsec$, i.e., within the ellipse-error of the \emph{IRAS} coordinates, 
with no preferential direction of the distance vector.  
In a few cases, the offset between the \emph{IRAS} and 2MASS coordinates 
is larger than the \emph{IRAS} coordinates uncertainty.  
Obviously, these are sources that have been identified through their 
GLIMPSE \emph{Spitzer} or \emph{MSX} images.  
A close inspection to the original \emph{IRAS} images reveals that these 
sources are either extended sources or mutiple, unresolved sources, thus 
introducing a significant offset in their \emph{IRAS} coordinates.

\onlfig{3}{
\begin{figure*}[!t]
\centering

\caption{
({\it left}) 
2MASS $J$ (blue), $H$ (green), and $K_{\rm S}$ (red) composite pictures and 
({\it right}) 
IR spectral energy distributions (SEDs) of the \emph{IRAS} post-AGB star and 
PN candidates with near-IR counterparts in the 2MASS PSC.  
In the pictures, north is up, east to the left, and the 
locations of the near-IR counterparts and their type 
(\emph{o} or \emph{n} whether they are detected in DSS 
red images or not) are overlaid.  
The SEDs are built using data extracted from the 2MASS, \emph{Spitzer} 
GLIMPSE, \emph{MSX}, and \emph{IRAS} catalogues. 
The arrows at 100 $\mu$m represent flux upper limits (IRAS quality factor, 
FQUAL, equals to 1).
}
\label{2MASSrgb1}
\end{figure*}

}

\onlfig{4}{
\begin{figure*}[!t]
\centering

\caption{
2MASS $J$ (blue), $H$ (green), and $K_{\rm S}$ (red) composite pictures of 
the \emph{IRAS} post-AGB star and PN candidates with near-IR counterparts 
that are not resolved or clearly detected in the 2MASS PSC.  
The locations of the near-IR counterparts and their type 
(\emph{o} or \emph{n} whether they are detected in DSS 
red images or not) are overlaid on the pictures.  
North is up, east to the left.  
}
\label{2MASSrgb2}
\end{figure*}

}

The sample of the 39 \emph{IRAS} post-AGB and PN candidates in Tab.~1 
with a 2MASS counterpart but dubious assignation in the 2MASS PSC are 
listed in Table~3 and their identification charts are shown in 
Figure.~\ref{2MASSrgb2}.

\begin{figure*}[!t]
\centering
\includegraphics[width=0.95\columnwidth]{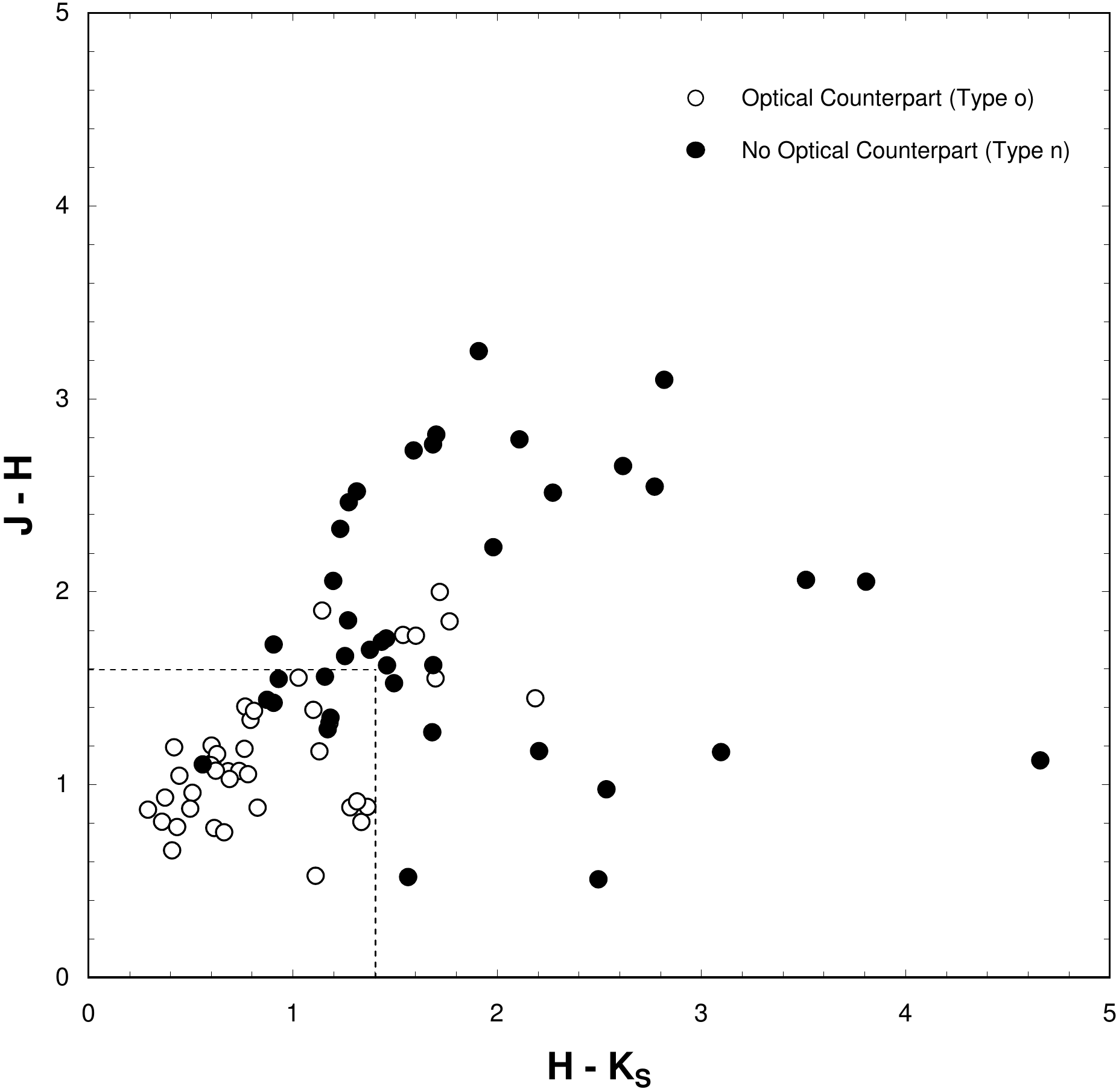}
\includegraphics[width=0.95\columnwidth]{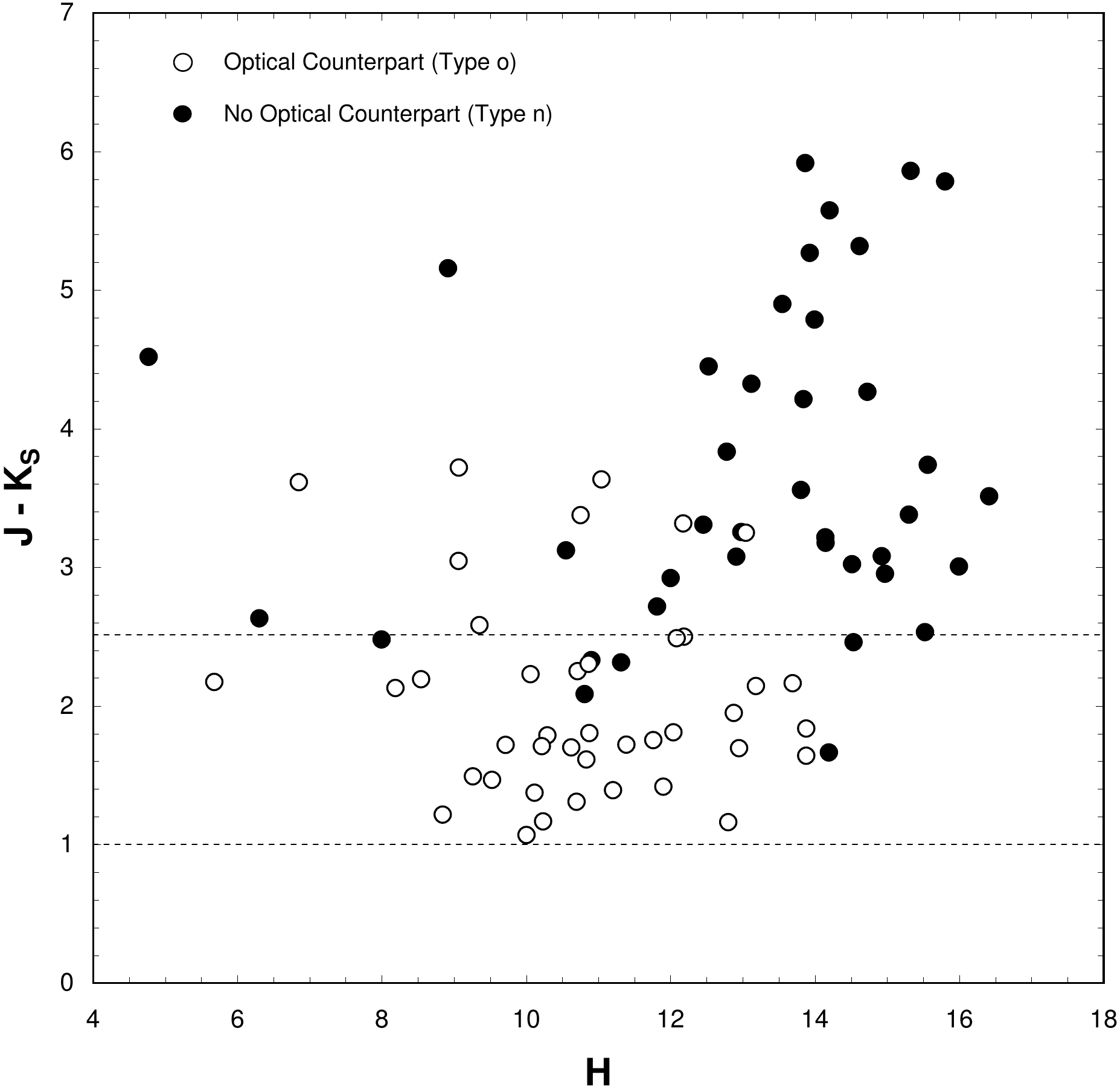}
\caption{
{\it (left)} $J-H$ vs.\ $H-K_s$ and {\it (right)} $J-K_s$ vs.\ $H$ 2MASS 
diagrams of the \emph{IRAS} post-AGB star and PN candidates in our sample 
with 2MASS PSC counterparts.  
The dashed lines mark the regions mostly occupied by sources of type 
\emph{o} (sources with optical counterpart).  
}
\label{2MASS}
\end{figure*}

Finally, the 46 sources that do not have a 2MASS 
counterpart are listed in Table~4.  
Most of these sources have been observed by \emph{MSX} and many of them 
have even \emph{Spitzer} GLIMPSE observations, but no near-IR counterpart 
is detected in 2MASS at the improved astrometric positions provided by 
the \emph{MSX} and \emph{Spitzer} GLIMPSE catalogues.  
Only IRAS\,16245$-$3859, IRAS\,18385$+$1350, and IRAS\,21537$+$6435 
have no \emph{Spitzer} GLIMPSE nor \emph{MSX} observations, and they 
do not have an obvious near-IR counterpart in 2MASS images within 
their \emph{IRAS} ellipse-error.  
The study of sources in Tab.~3 and 4 requires near-IR observations 
of greater spatial resolution and sensitivity, respectively, than 
the 2MASS observations used in this paper.  
These observations will be presented in a subsequent paper.

\addtocounter{table}{1}

\section{Discussion}

\subsection{Degree of Obscuration Among the \emph{IRAS} Sources}

Using the improved accuracy of the 2MASS coordinates, we have searched for 
the optical counterpart of the sources in Tab.~2 and 3 in the red plates 
of the Digitized Sky Survey\footnote{
The Digitized Sky Survey is based on photographic data obtained using the 
UK Schmidt Telescope and the Oschin Schmidt Telescope on Palomar Mountain. 
The UK Schmidt was operate by the Royal Observatory of Edinburgh, with 
funding from the UK Science and Engineering Research Council, until 1988 
June, and thereafter by the Anglo-Australian Observatory. 
The Palomar Observatory Sky Survey was funded by the National Geographic 
Society. 
The Oschin Schmidt Telescope is operated by the California Institute of 
Technology and Palomar Observatory. 
The plates were processed into the present compressed digital form with 
the permission of these institutes. 
The Digitized Sky Survey was produced at the Space Telescope Science 
Institute under US government grant NAGW-2166.} (DSS).  
The information on the optical detection of these sources, added 
to Tab.~2 and 3, unveils a significant population of optical 
counterparts for the a priori supposedly heavily obscured \emph{IRAS} 
post-AGB and PN candidates.  
Out of the 119 objects with 2MASS counterpart, 59  
have an optical counterpart, i.e., only 50\% of 
the objects in these two tables can be considered 
optically obscured.

Despite the limited spatial resolution and sensitivity of the 2MASS 
images and, very notably, of the DSS red images, the homogeneity of 
these two databases grants a brightness limited search that is 
meaningful for a statistical analysis.  
Therefore, the simple detection of an \emph{IRAS} source in our sample 
in the DSS and/or 2MASS images allows us to determine the degree of 
obscuration of this source.  
Accordingly, we can define three types of sources based on the shortest 
wavelength at which they are detected: 
\begin{itemize}
\item
Type \emph{m} sources (for mid-IR):  
Heavily obscured sources that are neither detected in DSS red 
images, nor in the 2MASS images.  
\item
Type \emph{n}  sources (for near-IR):  
Optically obscured sources detected in 2MASS images, but not 
in DSS images.  
\item
Type \emph{o} sources (for optical):  
Sources with little optical and near-IR obscuration detected 
both in DSS and 2MASS images.   
\end{itemize}

Type \emph{m} sources will be found among the sources listed in Tab.~4.  
The study of these sources will be postponed to an upcoming paper.  
Similarly, the 2MASS spatial resolution has not allowed us to measure 
the $JHK$ magnitudes of sources in Tab.~3 and the study of this sources 
will also be deferred.   
In the following, we will focus on the sources listed in Tab.~2, 
i.e., the sources of types \emph{n} and \emph{o} with a near-IR 
counterpart in the 2MASS PSC.

\begin{figure*}[!t]
\centering
\includegraphics[width=0.95\columnwidth]{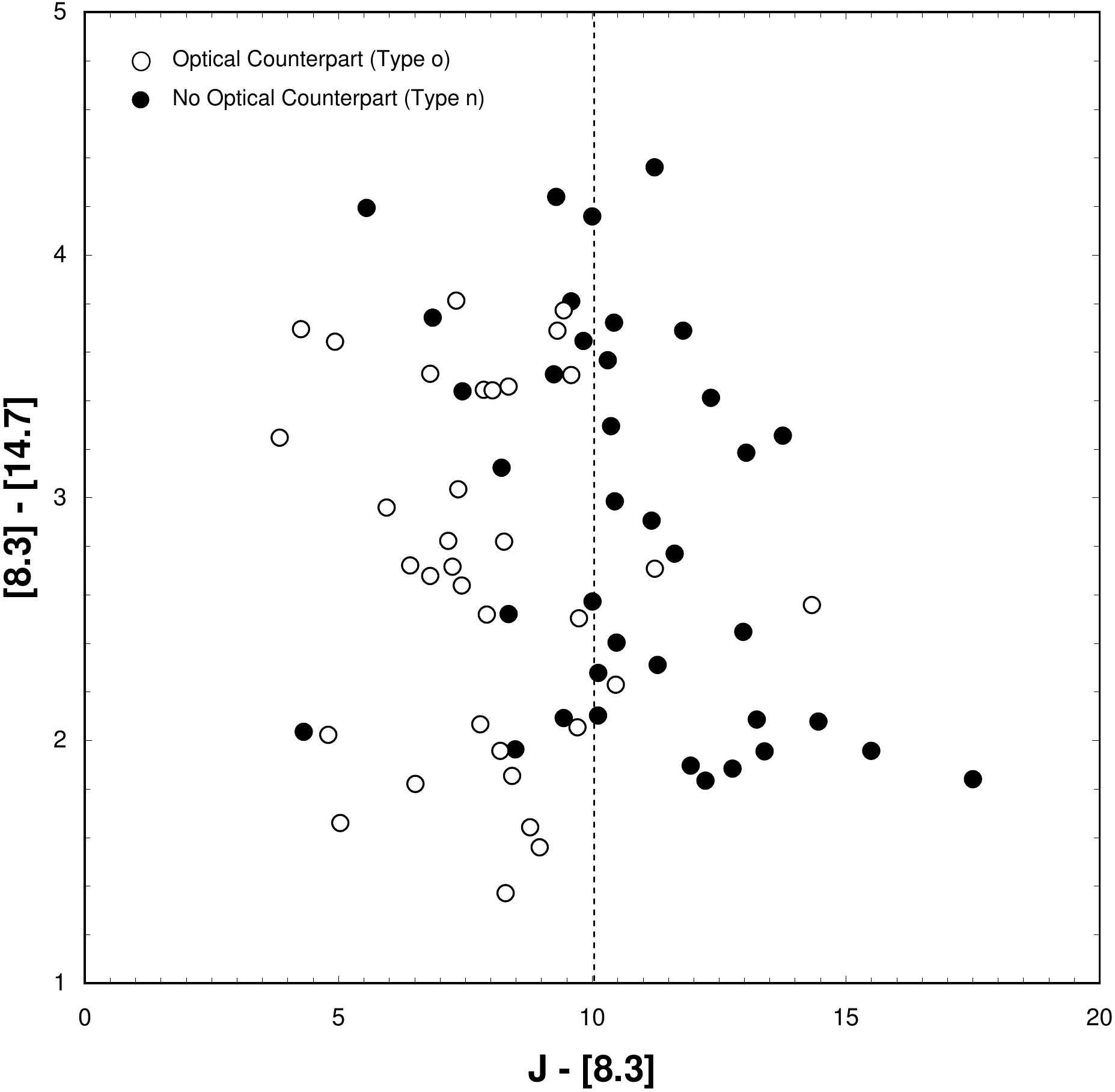}
\includegraphics[width=0.95\columnwidth]{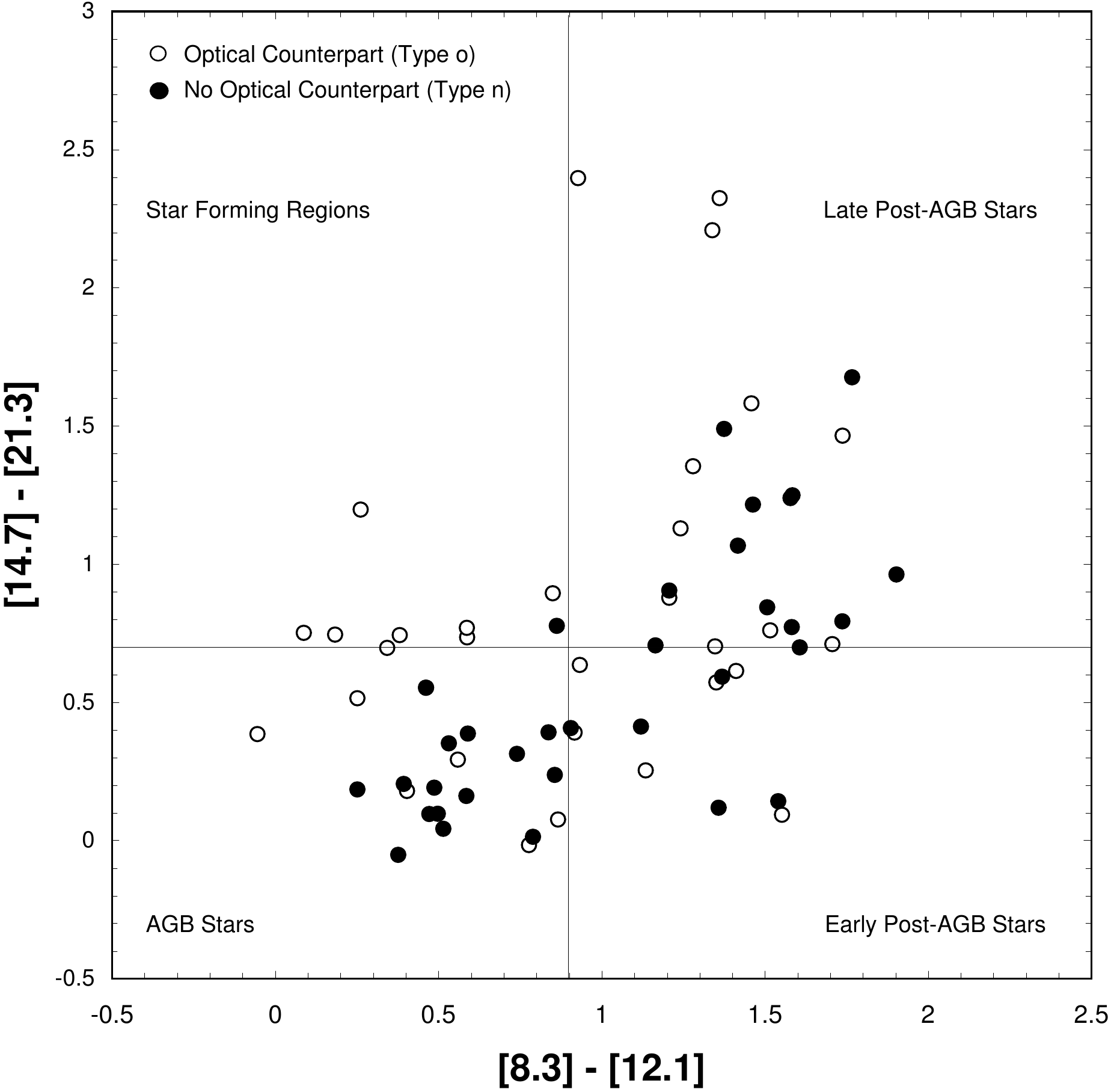}
\caption{
{\it (left)} \emph{MSX}--$J$ and {\it (right)} \emph{MSX} colour-colour 
diagrams of the \emph{IRAS} post-AGB star and PN candidates with 2MASS 
PSC and \emph{MSX} counterparts.  
The vertical line in the \emph{MSX}--$J$ colour-colour diagram at 
$J-[8.3]=10$ marks a rough division between sources of type \emph{o} 
and \emph{n}.  
The vertical and horizontal lines in the \emph{MSX} colour-colour 
diagram divide it into four quadrants that correspond to areas of 
higher probability of containing the different types of objects 
labeled on the diagram \citep{S02}.  
}
\label{MSX}
\end{figure*}

\subsection{Spectral Properties}

\begin{figure*}[!t]
\centering
\includegraphics[width=0.95\columnwidth]{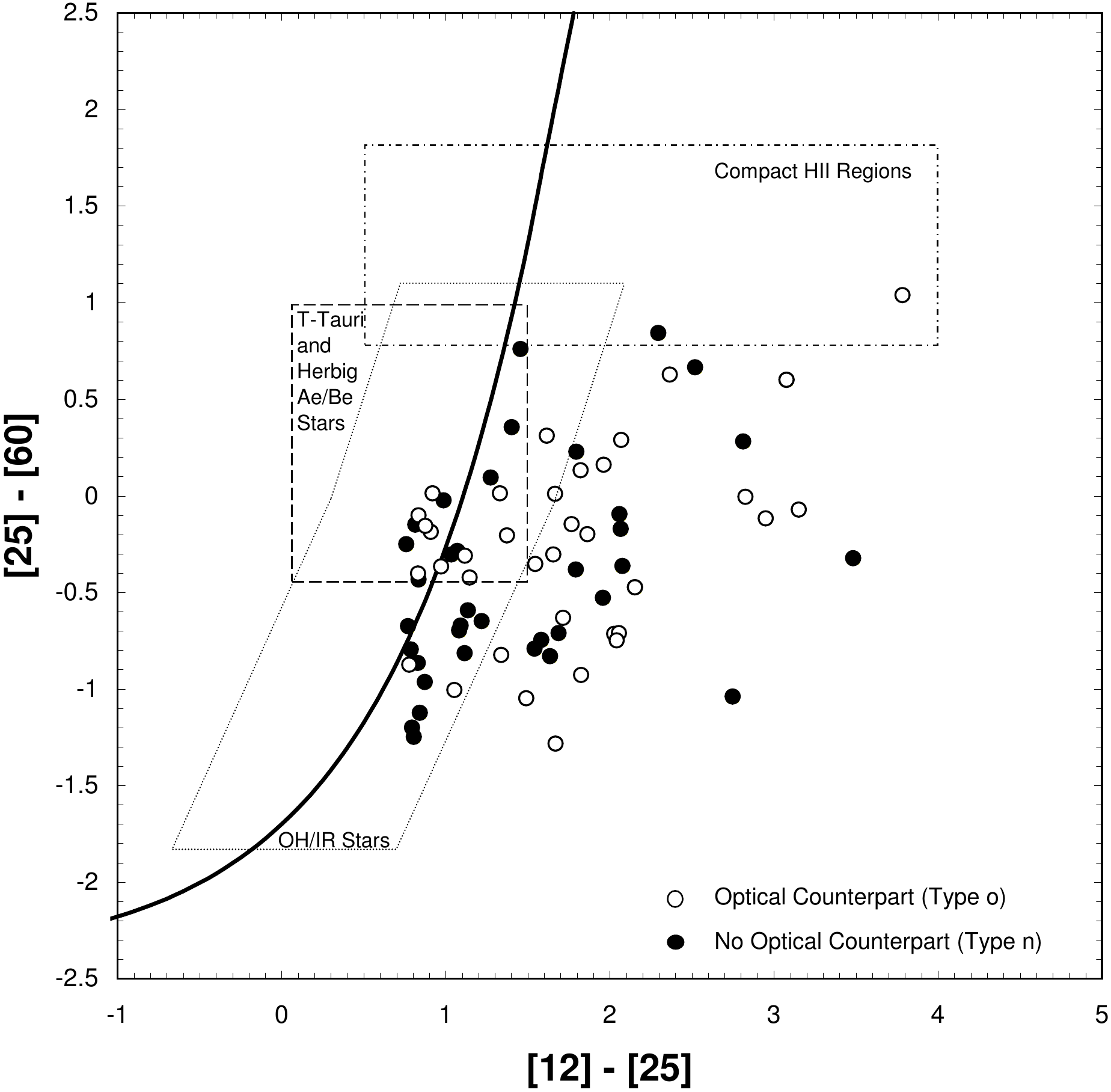}
\includegraphics[width=0.95\columnwidth]{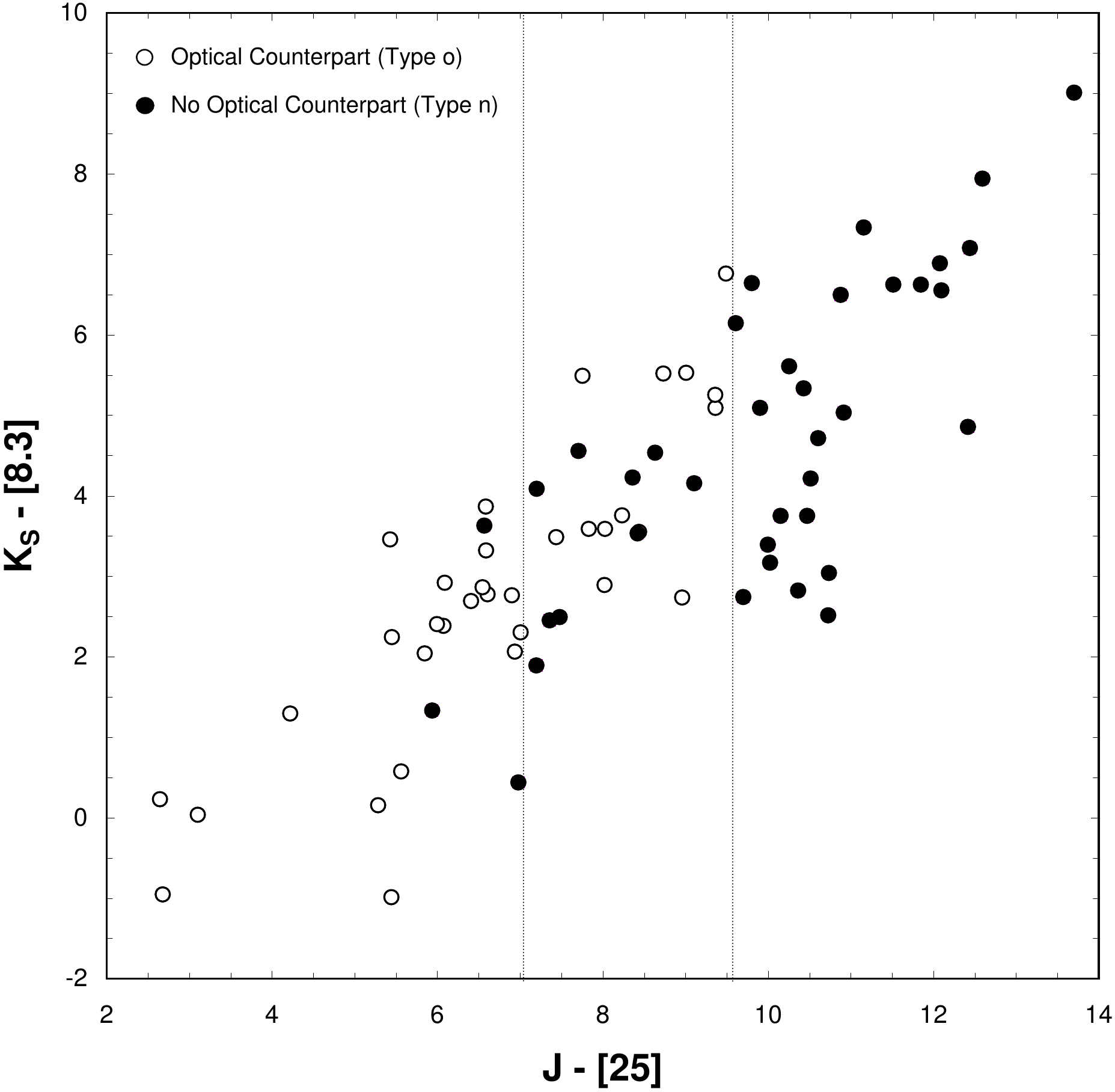}
\caption{
{\it (left)} \emph{IRAS} and {\it (right)} $K_s$--[8.3] vs. $J$--[25] 
colour-colour diagrams of the \emph{IRAS} post-AGB star and PN 
candidates with 2MASS PSC and \emph{MSX} counterparts.  
In the \emph{IRAS} colour-colour diagram, the locii of OH/IR stars, 
T-Tauri and Herbig Ae/Be stars, and compact H~{\sc ii} are overlaid 
as in Fig~\ref{IRAS}.  
The vertical lines in the $K_s$--[8.3] vs. $J$--[25] diagram mark 
three regions with the leftmost being dominated by sources with optical 
counterpart (type \emph{o} sources), and the rightmost being exclusively 
populated by sources with no optical counterpart (type \emph{n} sources).  
}
\label{IRAS_MSX_2MASS}
\end{figure*}

The near- and mid-IR properties of the type \emph{o} and type \emph{n} 
sources can be used to investigate their nature.  
Using the $JHK$ magnitudes of the sources listed in Tab.~2, the 
\emph{MSX} and \emph{Spitzer} GLIMPSE fluxes given in Table~5, 
and their \emph{IRAS} fluxes (Tab.~1), we have built the IR 
spectral energy distributions (SEDs) that are shown in 
Fig.~\ref{2MASSrgb1}.  
The analysis of the IR SEDs of these sources can be performed 
by studying the different colour diagrams shown in 
Figures~\ref{2MASS}, \ref{MSX}, and \ref{IRAS_MSX_2MASS}.

The $J-H$ vs.\ $H-K_s$ colour diagram (Fig.~\ref{2MASS}) shows that the 
sources in this sample 
have near-IR colours mostly in the ranges $0.5<J-H<3$ and $0.2<H-K_s<3$, 
although a few sources are very highly obscured, with $H-K_s$ colours up to 
$\sim$5 mag.  
In this diagram, sources detected in the optical tend to have 
smaller values of their $J-H$ and $H-K_s$ colours, with almost 
all these sources having $J-H<1.6$ and $H-K_s<1.4$.  
This trend is more clearly seen in the $J-K_s$ vs.\ $H$ diagram 
(Fig.~\ref{2MASS}) that shows 
noticeable differences in the distribution of sources of types $o$ and $n$; 
type $o$ sources are mostly located on the band $1<J-K_s<2.5$, 
while type $n$ sources are located mostly above this band, with 
$J-K_s\geq$2.5.  
The interpretation of this behaviour is clearly illustrated by 
the SED of these two types of sources:  
type \emph{o} sources use to have flat or shallow near-IR SEDs (e.g., 
IRAS\,14521$-$5300), while type \emph{n} sources usually present steep 
near-IR SEDs (e.g., IRAS\,14104$-$5819).

The segregation between the sources of type \emph{o} and \emph{n} 
with 2MASS PSC counterpart is also clear in the \emph{MSX}-$J$ 
colour diagram (Fig.~\ref{MSX}).  
In this diagram, sources of type \emph{o} have $J$--[8.3] colours 
typically lower than 10.  
On the contrary, sources of type \emph{o} and \emph{n} are not 
separated in the \emph{MSX} colour-colour diagram (Fig.~\ref{MSX}) 
on which both types are spread over a wide region.  
Interestingly, these sources are mostly located in the quadrants of this 
diagram that have been suggested to be populated by AGB and post-AGB stars, 
while generally avoiding the quadrant on which objects in star forming 
regions are typically found \citep{S02}.  
This result confirms our selection criteria for AGB and post-AGB stars 
in \S2.

Similarly, the \emph{IRAS} colour-colour diagram shown in 
Fig.~\ref{IRAS_MSX_2MASS} does not single out the optical 
type \emph{o} sources from the near-IR type \emph{n} sources 
in Tab.~2.  
On the other hand, the two types of sources show different 
properties in the $K_s$--[8.3] vs. $J$--[25] colour-colour 
diagram (Fig.~\ref{IRAS_MSX_2MASS}).  
In this diagram (spotlighted by Su\'arez et al., in prep., as able 
to discriminate very efficiently between different types of SEDs of 
post-AGB stars), sources of type \emph{o} have mostly $J$--[25]$<$7 and 
sources of type \emph{n} have $J$--[25]$>$9.6, with an intermediate 
region populated by both type \emph{o} and \emph{n} sources.  
It is interesting to note that the $K_s$--[8.3] and $J$--[25] colours 
of the sources in our sample are clearly correlated.

In all the colour-colour diagrams shown in Figs.~\ref{2MASS}, \ref{MSX}, 
and \ref{IRAS_MSX_2MASS}, the use of colours in which the $J$ magnitude 
is involved has allowed us to separate the sources with optical 
counterpart from these obscured in the optical.  
It can be concluded that the relation between the magnitude in the $J$ 
band and in different bands in the near-, mid-, and far-IR is critical 
in order to forecast the level of emission at optical wavelenghts.

\subsection{Sources in the \emph{Spitzer} GLIMPSE Image Atlas}

The \emph{Spitzer} images of the \emph{IRAS} post-AGB star and PN 
candidates with 
2MASS PSC counterparts that are available in the GLIMPSE Image Atlas 
are presented in Figure~\ref{Spitzer}.  
All \emph{IRAS} sources identified in the \emph{Spitzer} GLIMPSE Image 
Atlas are point-like, except IRAS\,18229$-$1127, IRAS\,18454$+$0001, 
and IRAS\,18576$+$0341, whose PSFs have FWHMs broader than those of 
point sources in their respective images.  
IRAS\,18229$-$1127 has a FWHM of 3\farcs0 in the 4 IRAC bands, 
just above the FWHM of stellar sources ($\sim$2\farcs1).  
IRAS\,18454$+$0001 is a point-like source, except in the 8~$\mu$m band, 
with a FWHM of 2\farcs4 (compared to the FWHM of 1\farcs9 of stellar 
sources).  
Finally, IRAS\,18576$+$0341 has the broadest FWHM, $\sim$5\farcs0  
(the FWHM of stellar sources is $\sim$2\farcs0), in the 3.6~$\mu$m, 
4.5~$\mu$m, and 5.8~$\mu$m bands (its image in the 8~$\mu$m band is 
saturated).

\onlfig{8}{
\begin{figure*}[!t]
\centering

\caption{
\emph{Spitzer} GLIMPSE composite pictures of the \emph{IRAS} post-AGB 
star and PN candidates with near-IR counterparts in the 2MASS PSC and 
\emph{Spitzer} observations.  
The blue, green, and red colours in the composite pictures correspond 
to the 3.6 $\mu$m, 4.5 $\mu$m, and 8.0 $\mu$m bands, respectively, 
except for the pictures of IRAS\,13421$-$6125, IRAS\,15408$-$5413, 
IRAS\,19374$+$2359, and IRAS\,18576$+$0341 where they represent the 
3.6 $\mu$m, 4.5 $\mu$m, and 5.8 $\mu$m bands, respectively, and for 
the picture of IRAS\,17548$-$2753 where the 4.5 $\mu$m band is shown 
in blue and green, and the 8.0 $\mu$m band is shown in red.  
The locations of the sources are overlaid on the pictures.  
North is up, east to the left.  
}
\label{Spitzer}
\end{figure*}

}

\section{Summary}

Using 2MASS, \emph{Spitzer} GLIMPSE, and \emph{MSX} data products, 
we have identified 119 near-IR counterparts out of a sample of 165 
\emph{IRAS} post-AGB star and PN candidates.  
For these sources, we provide near-IR identification charts.  
Among these 119 sources with near-IR counterpart, 80 objects have 
unambiguous 2MASS PSC counterpart.  
For these sources, we further provide improved coordinates, and IR 
energy spectral distributions from 1~$\mu$m to 100~$\mu$m.

Using the improved coordinates of the sources with 2MASS PSC counterpart, 
we have searched for their optical counterparts in DSS red plates and 
found that $\sim$50\% of these sources are detected in the optical.  
We have then investigated the different spectral properties of the 
sources with and without optical counterpart, dubbed sources of type 
\emph{o} and \emph{n}, respectively.  
The two types of sources have similar mid- and far-IR colours, 
but they appear clearly segregated in colour-colour diagrams 
in which the magnitude in the $J$ band is used to compute one 
of the colours.

\begin{acknowledgements}
      
This publication makes use of data products from the 2MASS, which is a 
joint project of the University of Massachusetts and the Infrared 
Processing and Analysis Center$/$California Institute of Technology, 
funded by the National Aeronautics and Space Administration and the 
National Science Foundation.  

This work is based, in part, on observations made with the \emph{Spitzer} 
Space Telescope, which is operated by the Jet Propulsion Laboratory, 
California Institute of Technology under a contract with NASA.  
Support for this work was provided by an award issued by JPL$/$Caltech.  
GRL acknowledges support of a postdoctoral fellowship from CONACyT 
(Mexico) grant 75861.  

MAG, OS, LFM, and JFG acknowledge support from Consejer\'{\i}a de 
Innovaci\'on, Ciencia y Empresa of Junta de Andaluc\'\i a.  
GRL, MAG, and LFM are partially funded by grant AYA2005-01495 of the 
Spanish Ministerio de Educaci\'on y Ciencia (MEC).  
JFG is partially funded by grant 2005-08523-C03-03 of MEC.  
GRL, MAG, OS, and LFM are partially funded by grant AYA2008-01934 of the 
Spanish Ministerio de Ciencia e Innovaci\'on (MICINN).  
OS and JFG are partially funded by grant 2008-06189-C03-01 of MICINN.  

This research has made use of the SIMBAD database, operated at CDS, 
Strasbourg, France.

\end{acknowledgements}

\longtab{1}{
\scriptsize{
\begin{longtable}{lccrrrrrrrccccc} 


\caption{\label{table1} 
\scriptsize{IRAS Identifications, Positions and Fluxes of Post-AGB Star 
and PN Candidates. }
} \\
\hline\hline
 Name & RA      & DEC     & Major Axis & Minor Axis & PA    & $F_{12}$ & $F_{25}$ & $F_{60}$ & $F_{100}$ & FQUAL$^\dagger$ \\ 
    ~ & (J2000) & (J2000) & ($\arcsec$)~~~~ & ($\arcsec$)~~~~ & ($^\circ$) & [{\rm Jy}] & [{\rm Jy}] & [{\rm Jy}] & [{\rm Jy}]   & ~  \\ 
\hline
\endfirsthead
\caption{continued.}\\
\hline\hline

\hline
\endhead
\hline

\endfoot

IRAS\,00509$+$6623 & 00 54 07.7 & $+$66 40 13 & 15~~~~~ &  4~~~~~ &  38 &  3.12 &  8.25 &  3.27 &  4.59 & 3 3 3 1 \\ 
IRAS\,04137$+$7016 & 04 19 08.8 & $+$70 23 23 & 15~~~~~ &  5~~~~~ &  75 &  0.65 &  2.12 &  1.39 &  1.12 & 3 3 3 1 \\ 
IRAS\,05573$+$3156 & 06 00 33.4 & $+$31 56 43 & 15~~~~~ &  5~~~~~ &  89 &  7.95 & 47.93 & 80.21 &108.90 & 3 3 3 3 \\ 
IRAS\,06499$+$0145 & 06 52 29.9 & $+$01 42 00 & 17~~~~~ &  5~~~~~ &  96 &  0.93 & 15.16 & 27.60 & 46.16 & 3 3 3 3 \\ 
IRAS\,08242$-$3828 & 08 26 03.5 & $-$38 38 48 & 29~~~~~ &  6~~~~~ & 116 & 11.58 & 26.77 & 22.53 & 10.75 & 3 3 3 3 \\ 
IRAS\,08351$-$4634 & 08 36 45.8 & $-$46 44 46 & 14~~~~~ &  4~~~~~ & 117 & 15.02 & 31.02 & 12.78 &  8.81 & 3 3 3 2 \\ 
IRAS\,09055$-$4629 & 09 07 19.4 & $-$46 41 21 & 15~~~~~ &  4~~~~~ & 126 &  0.74 &  3.05 &  1.47 & 32.70 & 3 3 3 1 \\ 
IRAS\,09102$-$5101 & 09 11 53.5 & $-$51 13 51 & 12~~~~~ &  4~~~~~ & 129 &  0.56 &  3.73 &  4.87 & 35.11 & 2 3 3 1 \\ 
IRAS\,09119$-$5150 & 09 13 32.9 & $-$52 02 39 & 14~~~~~ &  3~~~~~ & 129 &  0.79 &  3.82 &  2.13 & 25.94 & 3 3 2 1 \\ 
IRAS\,09370$-$4826 & 09 38 53.3 & $-$48 40 10 & 24~~~~~ &  5~~~~~ & 124 & 10.82 & 30.14 & 14.16 &  5.41 & 3 3 3 3 \\ 
IRAS\,09378$-$5117 & 09 39 36.9 & $-$51 31 26 & 15~~~~~ &  3~~~~~ & 136 &  1.87 &  6.43 &  3.01 & 16.31 & 3 3 3 1 \\ 
IRAS\,09500$-$5236 & 09 51 49.5 & $-$52 50 40 & 14~~~~~ &  3~~~~~ & 147 &  0.58 &  2.23 &  4.49 & 18.44 & 3 3 3 1 \\ 
IRAS\,10194$-$5625 & 10 21 15.2 & $-$56 40 32 & 17~~~~~ &  5~~~~~ & 146 & 16.75 & 57.02 & 27.78 & 52.91 & 3 3 3 1 \\ 
IRAS\,11339$-$6004 & 11 36 21.0 & $-$60 20 56 & 18~~~~~ &  3~~~~~ & 143 &  1.96 &  8.45 &  4.24 & 23.02 & 3 3 3 1 \\ 
IRAS\,11381$-$6401 & 11 40 32.3 & $-$64 18 33 & 14~~~~~ &  3~~~~~ & 146 &  5.41 & 27.76 & 17.02 & 72.42 & 3 3 3 1 \\ 
IRAS\,11444$-$6150 & 11 46 54.0 & $-$62 07 09 & 14~~~~~ &  3~~~~~ & 157 &  3.22 & 14.33 & 13.42 & 76.35 & 2 3 3 1 \\ 
IRAS\,11488$-$6432 & 11 51 17.0 & $-$64 49 09 & 15~~~~~ &  5~~~~~ & 144 &  1.56 &  3.37 &  3.07 & 14.49 & 3 3 3 1 \\ 
IRAS\,11544$-$6408 & 11 56 57.2 & $-$64 25 11 & 15~~~~~ &  4~~~~~ & 142 & 12.49 & 29.81 & 12.52 & 12.57 & 3 3 3 1 \\ 
IRAS\,11549$-$6225 & 11 57 30.8 & $-$62 42 12 & 12~~~~~ &  3~~~~~ & 149 &  7.60 & 20.06 & 10.79 & 33.30 & 3 3 3 3 \\ 
IRAS\,12262$-$6417 & 12 29 04.2 & $-$64 33 37 & 16~~~~~ &  6~~~~~ & 146 &  2.53 &  8.98 &  7.43 & 16.21 & 3 3 3 1 \\ 
IRAS\,12309$-$5928 & 12 33 44.8 & $-$59 45 18 & 15~~~~~ &  4~~~~~ & 139 &  1.99 &  5.44 &  2.93 & 19.26 & 3 3 3 1 \\ 
IRAS\,12360$-$5740 & 12 38 53.2 & $-$57 56 31 & 21~~~~~ &  6~~~~~ & 129 &  0.53 &  3.62 &  2.43 & 14.92 & 2 3 3 1 \\ 
IRAS\,12405$-$6219 & 12 43 31.1 & $-$62 36 13 & 16~~~~~ &  4~~~~~ & 132 & 16.63 &109.30 &250.70 &429.60 & 3 3 3 2 \\ 
IRAS\,13293$-$6000 & 13 32 38.4 & $-$60 15 36 & 16~~~~~ &  5~~~~~ & 129 &  0.75 &  1.95 &  4.99 & 34.31 & 3 3 3 1 \\ 
IRAS\,13356$-$6249 & 13 39 05.8 & $-$63 04 44 & 18~~~~~ &  6~~~~~ & 129 &  6.06 &149.90 &111.30 &202.70 & 3 3 3 1 \\ 
IRAS\,13398$-$5951 & 13 43 12.2 & $-$60 07 01 & 21~~~~~ &  4~~~~~ & 125 &  2.43 &  5.43 &  2.24 & 31.43 & 3 3 3 1 \\ 
IRAS\,13404$-$6059 & 13 43 52.8 & $-$61 14 46 & 25~~~~~ &  3~~~~~ & 129 &  5.24 & 15.76 & 12.00 & 86.92 & 3 3 3 1 \\ 
IRAS\,13421$-$6125 & 13 45 34.9 & $-$61 40 07 & 20~~~~~ &  5~~~~~ & 128 & 15.21 & 31.17 & 13.93 &122.10 & 3 3 3 1 \\ 
IRAS\,13427$-$6531 & 13 46 25.7 & $-$65 46 24 & 19~~~~~ &  6~~~~~ & 131 &  0.53 &  6.28 &  4.15 &  9.94 & 3 3 3 1 \\ 
IRAS\,13483$-$5905 & 13 51 44.1 & $-$59 20 16 & 22~~~~~ &  6~~~~~ & 125 &  6.05 & 16.94 & 12.72 & 27.99 & 3 3 3 1 \\ 
IRAS\,13500$-$6106 & 13 53 34.5 & $-$61 20 52 & 17~~~~~ &  6~~~~~ & 127 &  1.81 &  8.28 &  6.74 &123.20 & 3 3 3 1 \\ 
IRAS\,13529$-$5934 & 13 56 24.8 & $-$59 48 55 & 25~~~~~ &  7~~~~~ & 125 &  1.37 & 10.36 &  9.41 & 48.06 & 3 3 3 1 \\ 
IRAS\,14104$-$5819 & 14 14 00.4 & $-$58 33 57 & 23~~~~~ &  7~~~~~ & 128 &  5.56 & 11.56 &  3.83 & 29.82 & 3 3 3 1 \\ 
IRAS\,14249$-$5310 & 14 28 24.9 & $-$53 24 06 & 21~~~~~ &  5~~~~~ & 117 & 11.82 & 29.00 & 23.20 &  9.63 & 3 3 3 3 \\ 
IRAS\,14521$-$5300 & 14 55 45.9 & $-$53 12 33 & 23~~~~~ &  5~~~~~ & 109 &  2.44 &  6.46 &  2.58 & 13.13 & 3 3 3 1 \\ 
IRAS\,15038$-$5533 & 15 07 35.4 & $-$55 44 54 & 22~~~~~ &  5~~~~~ & 110 &  1.22 &  3.40 &  1.61 & 84.62 & 3 3 3 1 \\ 
IRAS\,15103$-$5754 & 15 14 18.9 & $-$58 05 20 & 18~~~~~ &  3~~~~~ &   2 & 10.80 &101.50 &126.50 &103.40 & 3 3 3 2 \\ 
IRAS\,15144$-$5812 & 15 18 22.0 & $-$58 23 13 & 17~~~~~ &  4~~~~~ & 116 &  7.67 & 35.59 & 35.89 &149.50 & 3 3 3 1 \\ 
IRAS\,15229$-$5433 & 15 26 42.3 & $-$54 44 23 & 17~~~~~ &  5~~~~~ & 110 &  0.42 &  3.84 &  2.27 & 80.68 & 2 3 3 1 \\ 
IRAS\,15284$-$6026 & 15 32 37.1 & $-$60 37 04 & 23~~~~~ &  5~~~~~ & 109 &  2.10 &  9.97 & 10.88 & 32.46 & 3 3 3 1 \\ 
IRAS\,15408$-$5657 & 15 44 48.1 & $-$57 07 08 & 23~~~~~ &  7~~~~~ & 110 & 26.60 & 54.08 & 29.03 & 16.22 & 3 3 2 2 \\ 
IRAS\,15408$-$5413 & 15 44 39.9 & $-$54 23 05 & 29~~~~~ &  4~~~~~ & 107 &166.90 &350.60 &111.00 &154.20 & 3 3 3 1 \\ 
IRAS\,15452$-$5459 & 15 49 11.5 & $-$55 08 52 & 21~~~~~ &  7~~~~~ & 103 & 87.05 &242.70 &273.60 &401.10 & 3 3 3 1 \\ 
IRAS\,15531$-$5704 & 15 57 10.4 & $-$57 13 20 & 28~~~~~ &  6~~~~~ & 102 &  2.27 &  4.69 &  2.27 & 70.81 & 3 3 3 1 \\ 
IRAS\,15534$-$5422 & 15 57 20.4 & $-$54 30 40 & 25~~~~~ &  6~~~~~ & 106 &  1.06 &  8.27 & 12.41 &355.80 & 3 3 2 1 \\ 
IRAS\,16209$-$4714 & 16 24 34.0 & $-$47 21 29 & 21~~~~~ &  4~~~~~ &  96 &  1.17 & 18.87 & 20.68 & 63.87 & 3 3 3 1 \\ 
IRAS\,16228$-$5014 & 16 26 36.4 & $-$50 21 05 & 13~~~~~ &  5~~~~~ & 146 &  0.79 & 25.68 & 66.78 &159.60 & 3 3 3 1 \\ 
IRAS\,16245$-$3859 & 16 27 53.7 & $-$39 05 46 & 43~~~~~ &  9~~~~~ &  99 &  3.78 & 25.27 & 31.62 & 16.13 & 3 3 3 3 \\ 
IRAS\,16279$-$8158 & 16 37 52.1 & $-$82 04 49 & 11~~~~~ &  4~~~~~ & 102 &  4.21 & 10.31 &  5.70 &  2.35 & 3 3 3 3 \\ 
IRAS\,16296$-$4507 & 16 33 12.7 & $-$45 13 44 & 17~~~~~ &  3~~~~~ &  97 &  3.26 & 17.67 & 14.20 & 31.32 & 3 3 3 1 \\ 
IRAS\,16333$-$4807 & 16 37 06.1 & $-$48 13 42 & 17~~~~~ &  4~~~~~ &  97 &  9.33 & 42.98 & 89.26 &113.20 & 3 3 3 1 \\ 
IRAS\,16507$-$4810 & 16 54 31.0 & $-$48 15 21 & 33~~~~~ & 11~~~~~ &  99 &  1.08 & 12.56 &  8.96 & 13.48 & 3 2 2 2 \\ 
IRAS\,16517$-$3626 & 16 55 05.1 & $-$36 31 30 & 34~~~~~ &  5~~~~~ & 100 &  1.51 &  5.97 &  2.27 & 11.38 & 3 3 3 1 \\ 
IRAS\,16518$-$3425 & 16 55 08.4 & $-$34 30 10 & 24~~~~~ &  6~~~~~ & 120 &  1.02 &  7.98 &  1.66 &  5.16 & 3 3 3 1 \\ 
IRAS\,16558$-$3417 & 16 59 10.5 & $-$34 22 05 & 30~~~~~ &  6~~~~~ &  98 &  2.63 & 11.60 &  7.68 & 10.85 & 3 3 3 2 \\ 
IRAS\,16559$-$2957 & 16 59 08.1 & $-$30 01 40 & 19~~~~~ &  5~~~~~ &  98 &  9.17 & 32.37 & 16.38 &  4.19 & 3 3 3 1 \\ 
IRAS\,16567$-$3838 & 17 00 08.2 & $-$38 43 08 & 40~~~~~ &  5~~~~~ &  99 &  7.30 & 15.68 &  7.07 & 58.57 & 3 3 3 1 \\ 
IRAS\,16584$-$3710 & 17 01 52.5 & $-$37 14 57 & 38~~~~~ &  6~~~~~ &  99 &  0.91 &  6.14 &  4.39 & 25.12 & 3 3 3 1 \\ 
IRAS\,17009$-$4154 & 17 04 29.7 & $-$41 58 35 & 33~~~~~ &  6~~~~~ &  99 &  7.44 & 86.38 & 65.84 &260.00 & 3 3 3 1 \\ 
IRAS\,17010$-$3810 & 17 04 26.9 & $-$38 14 43 & 24~~~~~ &  6~~~~~ &  99 &  2.23 & 14.78 &  7.68 & 30.04 & 3 3 3 1 \\ 
IRAS\,17021$-$3109 & 17 05 23.2 & $-$31 13 18 & 37~~~~~ &  5~~~~~ &  97 &  3.29 & 11.97 & 16.59 &  7.88 & 3 3 3 3 \\ 
IRAS\,17021$-$3054 & 17 05 24.1 & $-$30 58 13 & 23~~~~~ &  5~~~~~ &  97 &  1.18 &  5.34 &  2.49 & 16.78 & 3 3 3 1 \\ 
IRAS\,17052$-$3245 & 17 08 32.7 & $-$32 49 43 & 35~~~~~ &  6~~~~~ &  97 &  3.13 & 10.79 & 12.02 & 10.12 & 3 3 3 1 \\ 
IRAS\,17067$-$3759 & 17 10 08.3 & $-$38 03 22 & 29~~~~~ &  6~~~~~ &  98 &  1.98 & 10.60 & 10.66 & 68.33 & 3 3 3 1 \\ 
IRAS\,17088$-$4221 & 17 12 22.6 & $-$42 25 13 & 26~~~~~ &  5~~~~~ &  98 & 42.70 &128.30 &106.80 & 36.90 & 3 3 3 3 \\ 
IRAS\,17097$-$3624 & 17 13 05.1 & $-$36 27 54 & 42~~~~~ &  6~~~~~ &  97 &  1.52 &  6.62 & 10.11 &227.40 & 3 3 3 1 \\ 
IRAS\,17149$-$3053 & 17 18 11.7 & $-$30 56 40 & 36~~~~~ &  6~~~~~ &  96 &  2.13 & 11.12 &  7.82 & 20.93 & 3 3 3 1 \\ 
IRAS\,17150$-$3224 & 17 18 19.9 & $-$32 27 20 & 32~~~~~ &  6~~~~~ &  96 & 57.92 &322.30 &268.30 & 82.41 & 3 3 3 3 \\ 
IRAS\,17153$-$3814 & 17 18 44.6 & $-$38 17 23 & 15~~~~~ &  4~~~~~ &  97 &  2.64 & 19.55 & 52.88 &436.50 & 3 3 3 1 \\ 
IRAS\,17158$-$4049 & 17 19 20.9 & $-$40 52 43 & 46~~~~~ &  6~~~~~ &  97 &  0.99 &  6.56 &  7.14 &295.40 & 3 3 3 1 \\ 
IRAS\,17168$-$3736 & 17 20 15.0 & $-$37 39 32 & 18~~~~~ &  5~~~~~ &  97 &  9.79 & 36.98 & 47.89 &280.60 & 3 3 3 1 \\ 
IRAS\,17175$-$2819 & 17 20 42.4 & $-$28 22 36 & 42~~~~~ &  5~~~~~ &  95 &  3.38 & 14.94 &  5.23 & 10.18 & 3 3 3 1 \\ 
IRAS\,17233$-$2602 & 17 26 28.7 & $-$26 04 58 & 27~~~~~ &  5~~~~~ &  94 &  3.69 & 10.61 &  7.19 &  7.93 & 3 3 3 1 \\ 
IRAS\,17234$-$4008 & 17 26 55.6 & $-$40 11 03 & 37~~~~~ &  6~~~~~ &  96 &  1.48 & 12.46 & 10.48 & 83.69 & 3 3 3 1 \\ 
IRAS\,17269$-$2235 & 17 29 58.2 & $-$22 37 34 & 26~~~~~ &  7~~~~~ &  93 &  0.68 & 14.89 & 22.26 &  8.85 & 2 3 3 3 \\ 
IRAS\,17291$-$2147 & 17 32 10.1 & $-$21 49 59 & 24~~~~~ &  6~~~~~ &  93 &  2.54 & 12.18 &  8.24 &  6.42 & 3 3 3 1 \\ 
IRAS\,17301$-$2538 & 17 33 13.2 & $-$25 40 24 & 20~~~~~ &  5~~~~~ &  93 &  0.78 &  7.80 &  4.29 & 33.13 & 3 3 3 1 \\ 
IRAS\,17348$-$2906 & 17 38 04.2 & $-$29 08 23 & 39~~~~~ &  9~~~~~ &  93 &  4.71 & 11.55 &  8.24 &305.50 & 3 3 3 1 \\ 
IRAS\,17359$-$2902 & 17 39 07.7 & $-$29 04 02 & 41~~~~~ &  9~~~~~ &  93 &  2.04 & 12.38 &  7.61 &259.60 & 3 3 2 1 \\ 
IRAS\,17360$-$2142 & 17 39 05.9 & $-$21 43 52 & 25~~~~~ &  7~~~~~ &  92 &  0.93 & 12.53 & 12.46 &  6.52 & 3 3 3 1 \\ 
IRAS\,17361$-$4159 & 17 39 43.8 & $-$42 00 41 & 25~~~~~ &  5~~~~~ &  95 &  2.58 &  6.73 &  5.54 & 24.89 & 3 3 3 1 \\ 
IRAS\,17376$-$3448 & 17 40 56.4 & $-$34 50 03 & 22~~~~~ &  5~~~~~ &  93 &  3.70 & 10.18 &  6.92 & 13.33 & 3 3 3 1 \\ 
IRAS\,17382$-$2531 & 17 41 20.1 & $-$25 32 53 & 38~~~~~ &  5~~~~~ &  92 &  3.03 &  7.64 &  5.31 & 11.46 & 3 3 3 1 \\ 
IRAS\,17385$-$3332 & 17 41 52.2 & $-$33 33 41 & 22~~~~~ &  6~~~~~ &  93 &  2.88 & 13.25 & 10.01 &238.40 & 3 3 3 1 \\ 
IRAS\,17385$-$2413 & 17 41 38.3 & $-$24 14 41 & 32~~~~~ &  6~~~~~ &  92 &  3.73 &  7.67 &  3.70 &  9.93 & 3 3 3 1 \\ 
IRAS\,17393$-$2727 & 17 42 32.2 & $-$27 28 28 & 40~~~~~ &  7~~~~~ &  92 &  1.83 & 17.83 & 36.85 & 85.69 & 2 3 3 1 \\ 
IRAS\,17404$-$2713 & 17 43 39.4 & $-$27 14 18 & 20~~~~~ &  6~~~~~ &  92 &  3.99 & 20.74 & 15.49 & 83.59 & 3 3 3 1 \\ 
IRAS\,17418$-$3335 & 17 45 10.7 & $-$33 36 13 & 18~~~~~ &  5~~~~~ &  94 &  1.83 & 14.63 & 12.06 &161.90 & 3 3 3 1 \\ 
IRAS\,17443$-$2949 & 17 47 35.2 & $-$29 50 56 & 34~~~~~ &  6~~~~~ &  93 & 15.77 & 39.39 & 34.47 &354.10 & 3 3 3 1 \\ 
IRAS\,17479$-$3032 & 17 51 12.5 & $-$30 33 44 & 33~~~~~ &  7~~~~~ &  92 &  2.63 & 13.02 & 16.51 &106.10 & 2 3 2 1 \\ 
IRAS\,17482$-$2501 & 17 51 22.3 & $-$25 01 50 & 27~~~~~ &  8~~~~~ &  91 &  1.34 &  5.18 & 11.53 &381.60 & 2 3 3 1 \\ 
IRAS\,17487$-$1922 & 17 51 44.7 & $-$19 23 42 & 31~~~~~ &  8~~~~~ &  91 &  2.16 & 20.25 &  7.21 & 11.81 & 3 3 3 1 \\ 
IRAS\,17499$-$3520 & 17 53 19.7 & $-$35 21 11 & 30~~~~~ &  6~~~~~ &  92 &  2.82 & 13.10 &  4.02 &  6.10 & 3 3 3 1 \\ 
IRAS\,17506$-$2955 & 17 53 49.3 & $-$29 55 34 & 21~~~~~ &  6~~~~~ &  91 &  1.62 &  6.29 &  6.61 &194.40 & 3 3 3 1 \\ 
IRAS\,17516$-$2525 & 17 54 43.5 & $-$25 26 27 & 29~~~~~ &  6~~~~~ &  91 & 51.58 &115.60 &100.10 &292.10 & 3 3 3 1 \\ 
IRAS\,17540$-$2753 & 17 57 14.1 & $-$27 54 16 & 16~~~~~ &  4~~~~~ &  91 &  1.39 & 14.19 & 26.16 &341.00 & 2 3 3 1 \\ 
IRAS\,17543$-$3102 & 17 57 34.2 & $-$31 03 00 & 27~~~~~ &  6~~~~~ &  91 &  2.76 & 21.87 & 24.44 & 71.27 & 3 3 3 1 \\ 
IRAS\,17548$-$2753 & 17 57 57.7 & $-$27 53 20 & 29~~~~~ &  5~~~~~ &  91 &  1.27 & 16.95 & 21.95 &174.90 & 3 3 3 1 \\ 
IRAS\,17550$-$2800 & 17 58 10.6 & $-$28 00 26 & 37~~~~~ &  5~~~~~ &  91 &  2.52 &  5.89 &  5.96 &203.40 & 3 3 3 1 \\ 
IRAS\,17550$-$2120 & 17 58 04.9 & $-$21 21 07 & 32~~~~~ &  5~~~~~ &  90 &  5.38 & 21.08 & 30.81 & 43.64 & 3 3 3 1 \\ 
IRAS\,17552$-$2030 & 17 58 16.3 & $-$20 30 22 & 37~~~~~ &  6~~~~~ &  90 &  1.22 &  4.62 &  3.93 & 29.08 & 3 3 3 1 \\ 
IRAS\,17560$-$2027 & 17 59 04.4 & $-$20 27 23 & 24~~~~~ &  6~~~~~ &  90 &  1.59 & 15.32 & 17.79 & 40.99 & 3 3 3 1 \\ 
IRAS\,17580$-$3111 & 18 01 19.7 & $-$31 11 23 & 38~~~~~ &  7~~~~~ &  91 &  3.24 & 15.32 &  7.96 & 15.19 & 3 3 3 1 \\ 
IRAS\,17582$-$2619 & 18 01 21.2 & $-$26 19 37 & 24~~~~~ &  8~~~~~ &  90 &  1.38 &  9.28 &  7.92 &293.30 & 3 3 3 1 \\ 
IRAS\,17596$-$3952 & 18 03 06.2 & $-$39 51 55 & 34~~~~~ & 10~~~~~ &  92 &  0.48 &  1.11 &  0.59 & 12.68 & 3 2 2 1 \\ 
IRAS\,18011$-$1847 & 18 04 02.7 & $-$18 47 10 & 32~~~~~ &  5~~~~~ &  89 &  2.54 & 13.67 & 16.30 & 60.24 & 3 3 3 1 \\ 
IRAS\,18015$-$1352 & 18 04 22.2 & $-$13 51 49 & 29~~~~~ &  7~~~~~ &  89 &  2.11 &  4.24 &  3.37 & 21.74 & 3 3 3 2 \\ 
IRAS\,18016$-$2743 & 18 04 45.8 & $-$27 43 11 & 24~~~~~ &  5~~~~~ &  90 &  3.02 &  7.15 &  3.07 &131.80 & 3 3 3 1 \\ 
IRAS\,18039$-$1903 & 18 06 53.3 & $-$19 03 09 & 37~~~~~ &  6~~~~~ &  89 &  4.17 & 11.11 &  7.31 &261.90 & 3 3 2 1 \\ 
IRAS\,18049$-$2118 & 18 07 54.8 & $-$21 18 09 & 42~~~~~ &  6~~~~~ &  89 & 11.83 & 25.70 &  9.12 &206.50 & 3 3 3 1 \\ 
IRAS\,18051$-$2415 & 18 08 12.8 & $-$24 14 36 & 38~~~~~ &  6~~~~~ &  90 &  1.88 &  8.14 &  9.49 &172.30 & 3 3 3 1 \\ 
IRAS\,18071$-$1727 & 18 10 05.4 & $-$17 26 56 & 26~~~~~ &  7~~~~~ &  89 & 23.66 & 76.57 & 83.46 &304.70 & 3 3 3 1 \\ 
IRAS\,18083$-$2155 & 18 11 18.9 & $-$21 55 05 & 24~~~~~ &  6~~~~~ &  89 &  7.40 & 37.97 & 27.07 &496.00 & 3 3 3 1 \\ 
IRAS\,18087$-$1440 & 18 11 34.2 & $-$14 39 59 & 28~~~~~ &  7~~~~~ &  88 &  2.58 & 21.86 & 32.31 &217.20 & 3 3 3 1 \\ 
IRAS\,18105$-$1935 & 18 13 32.2 & $-$19 35 03 & 22~~~~~ &  7~~~~~ &  89 &  7.33 & 19.87 & 10.46 &369.50 & 3 3 3 1 \\ 
IRAS\,18113$-$2503 & 18 14 26.2 & $-$25 02 55 & 33~~~~~ &  6~~~~~ &  89 &  2.90 & 14.77 & 12.90 & 29.16 & 3 3 3 1 \\ 
IRAS\,18135$-$1456 & 18 16 25.6 & $-$14 55 15 & 20~~~~~ &  6~~~~~ &  88 & 31.02 &124.40 &157.60 &429.10 & 3 3 3 1 \\ 
IRAS\,18183$-$2538 & 18 21 24.7 & $-$25 36 35 & 22~~~~~ &  6~~~~~ &  88 &  1.64 &  3.91 &  2.05 & 33.40 & 3 3 3 1 \\ 
IRAS\,18199$-$1442 & 18 22 50.8 & $-$14 40 49 & 22~~~~~ &  6~~~~~ &  87 & 12.05 & 25.54 & 22.25 &282.60 & 3 3 2 1 \\ 
IRAS\,18229$-$1127 & 18 25 45.0 & $-$11 25 56 & 15~~~~~ &  6~~~~~ &  87 &  6.28 & 27.86 & 37.09 &133.50 & 3 2 3 2 \\ 
IRAS\,18236$-$0447 & 18 26 20.3 & $-$04 45 42 & 41~~~~~ &  6~~~~~ &  86 &  2.08 &  5.17 &  5.06 & 12.12 & 3 3 3 1 \\ 
IRAS\,18246$-$1032 & 18 27 24.0 & $-$10 30 24 & 30~~~~~ &  6~~~~~ &  86 &  2.19 & 20.25 & 50.37 &386.00 & 3 3 3 1 \\ 
IRAS\,18355$-$0712 & 18 38 15.4 & $-$07 09 52 & 31~~~~~ &  5~~~~~ &  85 &  1.73 & 14.34 & 31.14 &159.90 & 2 3 2 1 \\ 
IRAS\,18361$-$1203 & 18 38 58.8 & $-$12 00 44 & 37~~~~~ &  6~~~~~ &  86 &  2.24 &  4.83 &  3.33 & 52.79 & 3 3 3 1 \\ 
IRAS\,18385$+$1350 & 18 40 52.0 & $+$13 52 53 & 13~~~~~ &  4~~~~~ &  26 &  0.49 &  3.34 &  3.61 &  3.83 & 3 3 3 3 \\ 
IRAS\,18434$-$0042 & 18 46 04.4 & $-$00 38 55 & 30~~~~~ &  6~~~~~ &  83 &  1.97 &  5.61 &  3.25 &343.40 & 3 3 2 1 \\ 
IRAS\,18454$+$0001 & 18 48 01.5 & $+$00 04 47 & 39~~~~~ &  8~~~~~ &  83 &  0.80 & 14.54 & 13.61 &383.50 & 2 3 3 1 \\ 
IRAS\,18470$+$0015 & 18 49 39.1 & $+$00 18 52 & 38~~~~~ &  6~~~~~ &  83 &  4.11 & 10.67 &  8.07 &249.60 & 3 3 3 1 \\ 
IRAS\,18485$+$0642 & 18 50 58.9 & $+$06 45 55 & 18~~~~~ &  6~~~~~ &  80 &  3.58 & 21.86 & 25.32 & 61.08 & 3 3 3 1 \\ 
IRAS\,18514$+$0019 & 18 53 57.9 & $+$00 23 24 & 29~~~~~ &  6~~~~~ &  81 &  4.95 & 23.38 & 17.26 &152.20 & 3 3 3 1 \\ 
IRAS\,18524$+$0544 & 18 54 54.1 & $+$05 48 11 & 21~~~~~ &  5~~~~~ &  75 &  0.37 &  5.61 &  5.04 & 39.88 & 3 3 3 1 \\ 
IRAS\,18529$+$0210 & 18 55 26.3 & $+$02 14 49 & 22~~~~~ &  6~~~~~ &  81 &  5.69 & 16.24 & 15.24 &1921.00& 3 3 3 1 \\ 
IRAS\,18576$+$0341 & 19 00 10.9 & $+$03 45 47 & 23~~~~~ &  5~~~~~ &  81 & 58.48 &425.00 &274.70 &1660.00& 3 3 3 1 \\ 
IRAS\,18580$+$0818 & 19 00 25.2 & $+$08 22 46 & 27~~~~~ &  6~~~~~ &  79 &  0.90 &  3.35 &  2.31 & 16.39 & 3 3 3 1 \\ 
IRAS\,18596$+$0315 & 19 02 06.2 & $+$03 20 16 & 24~~~~~ &  8~~~~~ &  80 &  2.60 & 14.17 & 22.57 &112.80 & 3 3 3 1 \\ 
IRAS\,19006$+$1022 & 19 02 59.9 & $+$10 26 35 & 29~~~~~ &  8~~~~~ &  79 &  1.02 &  5.34 &  6.58 & 12.85 & 3 3 3 1 \\ 
IRAS\,19011$+$1049 & 19 03 30.7 & $+$10 53 53 & 50~~~~~ &  6~~~~~ &  79 &  9.40 & 26.10 & 12.33 &  9.26 & 3 3 3 1 \\ 
IRAS\,19013$+$0629 & 19 03 44.4 & $+$06 34 12 & 30~~~~~ & 12~~~~~ &  79 &  5.12 & 17.60 & 16.31 & 79.21 & 3 3 3 1 \\ 
IRAS\,19015$+$1256 & 19 03 52.6 & $+$13 01 21 & 23~~~~~ &  5~~~~~ &  38 &  0.92 &  4.20 &  1.33 &  7.60 & 3 3 2 1 \\ 
IRAS\,19071$+$0857 & 19 09 29.7 & $+$09 02 23 & 26~~~~~ &  9~~~~~ &  79 &  0.94 & 17.26 & 13.06 & 72.03 & 3 3 3 1 \\ 
IRAS\,19075$+$0432 & 19 10 00.0 & $+$04 37 06 & 27~~~~~ &  8~~~~~ &  80 &  5.25 & 28.13 & 31.77 & 14.36 & 3 3 3 3 \\ 
IRAS\,19079$-$0315 & 19 10 32.4 & $-$03 10 16 & 23~~~~~ &  7~~~~~ &  81 &  2.27 & 12.03 &  4.22 & 16.56 & 3 3 3 1 \\ 
IRAS\,19083$+$0119 & 19 10 54.4 & $+$01 24 42 & 21~~~~~ &  7~~~~~ &  80 &  2.34 & 15.58 & 14.27 &  4.72 & 3 3 3 2 \\ 
IRAS\,19094$+$1627 & 19 11 44.5 & $+$16 32 54 & 25~~~~~ &  7~~~~~ &  78 &  0.80 &  5.17 &  2.68 &  5.85 & 3 3 3 1 \\ 
IRAS\,19134$+$2131 & 19 15 35.4 & $+$21 36 33 & 32~~~~~ &  5~~~~~ &  77 &  5.06 & 15.56 &  8.56 &  3.95 & 3 3 3 1 \\ 
IRAS\,19176$+$1251 & 19 19 55.8 & $+$12 57 34 & 27~~~~~ &  6~~~~~ &  78 &  0.85 & 10.70 &  4.11 & 71.30 & 3 3 3 1 \\ 
IRAS\,19178$+$1206 & 19 20 14.0 & $+$12 12 20 & 17~~~~~ &  6~~~~~ &  78 &  3.66 &  8.57 &  4.85 & 68.21 & 3 3 3 1 \\ 
IRAS\,19181$+$1806 & 19 20 25.2 & $+$18 11 41 & 37~~~~~ &  5~~~~~ &  78 &  0.47 &  4.14 &  7.38 & 14.09 & 2 3 3 1 \\ 
IRAS\,19190$+$1102 & 19 21 25.3 & $+$11 08 40 & 28~~~~~ &  5~~~~~ &  78 &  1.59 & 13.67 & 24.52 & 20.41 & 3 3 3 3 \\ 
IRAS\,19193$+$1804 & 19 21 31.5 & $+$18 10 09 & 27~~~~~ &  6~~~~~ &  78 &  0.54 &  9.11 & 15.84 & 58.62 & 3 3 3 1 \\ 
IRAS\,19208$+$1541 & 19 23 06.8 & $+$15 47 33 & 26~~~~~ &  5~~~~~ &  78 &  0.77 &  1.57 &  3.37 & 92.78 & 3 3 3 1 \\ 
IRAS\,19315$+$2235 & 19 33 41.6 & $+$22 42 08 & 16~~~~~ &  5~~~~~ &  74 &  1.05 &  5.68 &  4.41 &  8.80 & 3 3 3 1 \\ 
IRAS\,19319$+$2214 & 19 34 03.5 & $+$22 21 16 & 24~~~~~ &  4~~~~~ &  74 &  3.18 &  6.58 &  3.16 & 39.29 & 3 3 3 1 \\ 
IRAS\,19374$+$2359 & 19 39 35.6 & $+$24 06 28 & 32~~~~~ &  5~~~~~ &  71 & 23.62 & 98.18 & 70.87 &768.40 & 3 3 3 1 \\ 
IRAS\,19454$+$2920 & 19 47 24.2 & $+$29 28 11 & 19~~~~~ &  5~~~~~ &  69 & 17.27 & 89.56 & 54.43 & 14.66 & 3 3 3 3 \\ 
IRAS\,20035$+$3242 & 20 05 29.6 & $+$32 51 35 & 20~~~~~ &  3~~~~~ &  66 &  7.49 & 25.59 & 21.16 & 16.39 & 3 3 3 2 \\ 
IRAS\,20042$+$3259 & 20 06 10.6 & $+$33 07 51 & 22~~~~~ &  5~~~~~ &  64 &  1.49 &  4.40 &  1.89 & 41.49 & 3 3 3 1 \\ 
IRAS\,20174$+$3222 & 20 19 28.0 & $+$32 32 13 & 21~~~~~ &  5~~~~~ &  64 &  2.44 & 16.02 &  8.04 &  8.73 & 3 3 3 2 \\ 
IRAS\,20214$+$3749 & 20 23 19.2 & $+$37 58 52 & 22~~~~~ &  5~~~~~ &  67 &  1.95 &  4.21 &  2.82 &184.10 & 3 3 2 1 \\ 
IRAS\,20244$+$3509 & 20 26 25.4 & $+$35 19 14 & 20~~~~~ &  6~~~~~ &  67 &  1.27 &  4.31 &  4.36 &179.30 & 3 3 3 1 \\ 
IRAS\,20406$+$2953 & 20 42 45.9 & $+$30 04 06 & 36~~~~~ &  5~~~~~ &  67 & 12.48 & 68.37 & 49.77 & 13.00 & 3 3 3 2 \\ 
IRAS\,20461$+$3853 & 20 48 04.6 & $+$39 05 00 & 31~~~~~ &  5~~~~~ &  57 &  2.10 & 11.28 &  4.80 &  4.46 & 3 3 3 1 \\ 
IRAS\,21525$+$5643 & 21 54 15.0 & $+$56 57 23 &  9~~~~~ &  3~~~~~ &  15 &  4.33 & 11.62 &  8.93 & 20.47 & 3 3 3 1 \\ 
IRAS\,21537$+$6435 & 21 55 04.5 & $+$64 49 54 & 12~~~~~ &  4~~~~~ & 161 &  6.91 & 26.10 & 13.34 &  6.07 & 3 3 3 1 \\ 
IRAS\,21554$+$6204 & 21 56 58.1 & $+$62 18 43 &  8~~~~~ &  4~~~~~ &   9 & 62.94 &138.20 & 52.13 & 15.59 & 3 3 3 3 \\ 

\end{longtable}
\vspace*{-0.2cm}
\noindent
$^\dagger$
The \emph{IRAS} flux quality factor, FQUAL, indicates whether the flux values 
are of high quality (3), moderate quality (2), or simply represent an upper 
limit of the flux (1).
The four values in this column refer consecutively to the flux quality in the 12 
$\mu$m, 25 $\mu$m, 60 $\mu$m, and 100 $\mu$m bands.
}
}


\begin{table*}
\scriptsize{
\caption{\footnotesize{
\emph{IRAS} Post-AGB Star and PN Candidates with Near-IR Counterparts in the 2MASS PSC. }}

\label{table:2}      
\centering          
\begin{tabular}{rccrrrlccc}     
\hline\hline       

\multicolumn{1}{c}{Object} & 
\multicolumn{1}{c}{$\alpha$} & 
\multicolumn{1}{c}{$\delta$} & 
\multicolumn{1}{c}{$\Delta\alpha$} & 
\multicolumn{1}{c}{$\Delta\delta$} & 
\multicolumn{1}{c}{Distance} & 
\multicolumn{1}{c}{Optical} & 
\multicolumn{1}{c}{$J$} & 
\multicolumn{1}{c}{$H$} & 
\multicolumn{1}{c}{$K$} \\
\multicolumn{1}{c}{} & 
\multicolumn{1}{c}{(J2000)} & 
\multicolumn{1}{c}{(J2000)} & 
\multicolumn{1}{c}{($\arcsec$)} & 
\multicolumn{1}{c}{($\arcsec$)} & 
\multicolumn{1}{c}{($\arcsec$)} & 
\multicolumn{1}{c}{} & 
\multicolumn{1}{c}{(mag)} & 
\multicolumn{1}{c}{(mag)} & 
\multicolumn{1}{c}{(mag)} \\

\hline

IRAS\,00509$+$6623 & 00 54 07.67 & $+$66 40 12.8 & $-$0.2 & $-$0.2 & 0.3~~~ & ~~~Yes & 13.950$\pm$0.032 & 12.175$\pm$0.032 & 10.634$\pm$0.024 \\
IRAS\,08242$-$3828 & 08 26 03.78 & $-$38 38 47.4 &    3.3 &    0.6 & 3.3~~~ & ~~~Yes & ~7.08$\pm$0.04   &  ~5.68$\pm$0.04  & ~4.912$\pm$0.022 \\
IRAS\,09055$-$4629 & 09 07 19.56 & $-$46 41 23.6 &    1.6 & $-$2.6 & 3.1~~~ & ~~~No  & 15.90$\pm$0.08   & 14.14$\pm$0.04   & 12.686$\pm$0.032 \\
IRAS\,09102$-$5101 & 09 11 57.36 & $-$51 14 24.9 &   36.3 &$-$33.9 &49.6~~~ & ~~~Yes & 14.41$\pm$0.05   & 13.88$\pm$0.06   & 12.77$\pm$0.05   \\
IRAS\,09119$-$5150 & 09 13 33.03 & $-$52 02 41.6 &    1.2 & $-$2.6 & 2.9~~~ & ~~~Yes & 13.99$\pm$0.04   & 13.18$\pm$0.04   & 11.85$\pm$0.04   \\
IRAS\,09378$-$5117 & 09 39 37.02 & $-$51 31 29.3 &    1.1 & $-$3.3 & 3.5~~~ & ~~~Yes & 12.077$\pm$0.026 & 10.876$\pm$0.022 & 10.273$\pm$0.021 \\
IRAS\,09500$-$5236 & 09 51 50.04 & $-$52 50 52.1 &    4.9 &$-$12.1 &13.1~~~ & ~~~No  & 17.2             & 14.62$\pm$0.06   & 11.845$\pm$0.023 \\
IRAS\,11339$-$6004 & 11 36 20.69 & $-$60 20 53.3 & $-$2.3 &    2.7 & 3.5~~~ & ~~~No  & 15.30$\pm$0.07   & 14.20$\pm$0.07   & 13.64$\pm$0.05   \\
IRAS\,11488$-$6432 & 11 51 17.36 & $-$64 49 12.7 &    2.3 & $-$3.7 & 4.4~~~ & ~~~Yes & 10.907$\pm$0.026 & ~9.354$\pm$0.024 & ~8.325$\pm$0.027 \\
IRAS\,12262$-$6417 & 12 29 04.49 & $-$64 33 38.0 &    1.9 & $-$1.0 & 2.1~~~ & ~~~Yes & 12.53$\pm$0.04   & 10.754$\pm$0.029 & ~9.150$\pm$0.023 \\
IRAS\,12309$-$5928 & 12 33 44.58 & $-$59 45 18.5 & $-$1.7 & $-$0.5 & 1.7~~~ & ~~~No  & 16.07$\pm$0.11   & 13.843$\pm$0.030 & 11.861$\pm$0.025 \\
IRAS\,13356$-$6249 & 13 39 06.31 & $-$63 04 44.3 &    3.5 & $-$0.3 & 3.5~~~ & ~~~No  & ~9.546$\pm$0.024 & ~8.00$\pm$0.06   & ~7.068$\pm$0.024 \\
IRAS\,13398$-$5951 & 13 43 12.56 & $-$60 07 03.5 &    2.7 & $-$2.5 & 3.7~~~ & ~~~No  & 13.372$\pm$0.027 & 11.813$\pm$0.023 & 10.655$\pm$0.021 \\
IRAS\,13421$-$6125 & 13 45 34.08 & $-$61 40 03.8 & $-$5.8 &    3.2 & 6.7~~~ & ~~~Yes & 12.027$\pm$0.026 & 10.835$\pm$0.022 & 10.415$\pm$0.023 \\
IRAS\,13483$-$5905 & 13 51 43.73 & $-$59 20 15.5 & $-$2.8 &    0.5 & 2.9~~~ & ~~~Yes & 13.669$\pm$0.021 & 12.800$\pm$0.022 & 12.508$\pm$0.021 \\
IRAS\,14104$-$5819 & 14 14 00.56 & $-$58 33 58.0 &    1.3 & $-$1.0 & 1.6~~~ & ~~~No  & 16.3             & 14.204$\pm$0.09  & 10.690$\pm$0.024 \\
IRAS\,14521$-$5300 & 14 55 45.70 & $-$53 12 30.1 & $-$1.8 &    2.9 & 3.4~~~ & ~~~Yes & 11.042$\pm$0.022 & 10.236$\pm$0.023 & ~9.876$\pm$0.023 \\
IRAS\,15038$-$5533 & 15 07 34.77 & $-$55 44 50.7 & $-$5.3 &    3.3 & 6.3~~~ & ~~~No  & 15.04$\pm$0.07   & 12.98$\pm$0.04   & 11.784$\pm$0.027 \\
IRAS\,15144$-$5812 & 15 18 21.92 & $-$58 23 11.9 & $-$0.6 &    1.1 & 1.3~~~ & ~~~Yes & 11.066$\pm$0.026 & ~9.068$\pm$0.023 & ~7.348$\pm$0.018 \\
IRAS\,15408$-$5657 & 15 44 48.36 & $-$57 07 08.7 &    2.1 & $-$0.7 & 2.2~~~ & ~~~No  & 17.4             & 15.3             & 11.519$\pm$0.026 \\
IRAS\,15408$-$5413 & 15 44 39.80 & $-$54 23 05.0 & $-$0.9 &    0.0 & 0.9~~~ & ~~~No  & ~7.583$\pm$0.021 & ~4.769$\pm$0.015 & ~3.07$\pm$0.31   \\
IRAS\,16228$-$5014 & 16 26 34.27 & $-$50 21 01.8 &$-$20.4 &    3.2 &20.6~~~ & ~~~Yes & ~9.927$\pm$0.026 & ~8.55$\pm$0.05   & ~7.735$\pm$0.023 \\
IRAS\,16279$-$8158 & 16 37 51.63 & $-$82 04 49.9 & $-$1.0 & $-$0.9 & 1.3~~~ & ~~~Yes & 10.308$\pm$0.023 & ~9.263$\pm$0.025 & ~8.818$\pm$0.023 \\
IRAS\,16517$-$3626 & 16 55 06.20 & $-$36 31 32.2 &   13.3 & $-$2.2 &13.4~~~ & ~~~Yes & 12.652$\pm$0.026 & 11.900$\pm$0.025 & 11.2             \\
IRAS\,16559$-$2957 & 16 59 08.22 & $-$30 01 40.3 &    1.6 & $-$0.3 & 1.6~~~ & ~~~Yes & 11.596$\pm$0.023 & 10.713$\pm$0.023 & ~9.347$\pm$0.021 \\
IRAS\,16567$-$3838 & 17 00 09.07 & $-$38 43 09.1 &   10.2 & $-$1.1 &10.2~~~ & ~~~No  & 15.9             & 14.7             & 11.632$\pm$0.027 \\
IRAS\,16584$-$3710 & 17 01 52.05 & $-$37 14 53.7 & $-$5.4 &    3.3 & 6.3~~~ & ~~~No  & 16.5             & 14.00$\pm$0.04   & 11.721$\pm$0.025 \\
IRAS\,17010$-$3810 & 17 04 27.31 & $-$38 14 41.7 &    4.8 &    1.3 & 5.0~~~ & ~~~Yes & 13.110$\pm$0.022 & 12.040$\pm$0.028 & 11.302$\pm$0.020 \\
IRAS\,17021$-$3109 & 17 05 23.71 & $-$31 13 18.7 &    6.5 & $-$0.7 & 6.6~~~ & ~~~No  & 17.4             & 16.4             & 13.88$\pm$0.05   \\
IRAS\,17021$-$3054 & 17 05 23.83 & $-$30 58 13.0 & $-$3.5 &    0.0 & 3.5~~~ & ~~~No  & 14.61$\pm$0.08   & 12.91$\pm$0.06   & 11.532$\pm$0.034 \\
IRAS\,17149$-$3053 & 17 18 10.93 & $-$30 56 39.9 & $-$9.9 &    0.1 & 9.9~~~ & ~~~No  & 15.8             & 14.5             & 13.37$\pm$0.06   \\
IRAS\,17175$-$2819 & 17 18 19.85 & $-$32 27 21.6 & $-$0.9 & $-$1.4 & 1.7~~~ & ~~~Yes & 11.10$\pm$0.04   & 10.22$\pm$0.04   & ~9.391$\pm$0.024 \\
IRAS\,17233$-$2602 & 17 26 28.99 & $-$26 04 56.8 &    3.9 &    1.2 & 4.1~~~ & ~~~Yes & 14.93$\pm$0.04   & 13.88$\pm$0.04   & 13.10$\pm$0.04   \\
IRAS\,17291$-$2147 & 17 32 11.17 & $-$21 50 02.2 &   14.9 & $-$3.2 &15.2~~~ & ~~~Yes & 13.507$\pm$0.021 & 12.188$\pm$0.022 & 11.008$\pm$0.019 \\
IRAS\,17348$-$2906 & 17 38 03.92 & $-$29 08 16.5 & $-$3.7 &    6.5 & 7.5~~~ & ~~~Yes & ~8.027$\pm$0.023 & ~6.302$\pm$0.047 & ~5.395$\pm$0.018 \\
IRAS\,17359$-$2902 & 17 39 07.70 & $-$29 04 02.8 & $-$0.1 &    0.1 & 0.2~~~ & ~~~No  & 12.75$\pm$0.06   & 11.32$\pm$0.07   & 10.44$\pm$0.04 \\
IRAS\,17360$-$2142 & 17 39 05.95 & $-$21 43 51.4 &    0.7 &    0.6 & 0.9~~~ & ~~~Yes & 10.825$\pm$0.023 & ~9.714$\pm$0.022 & ~9.107$\pm$0.019 \\
IRAS\,17385$-$3332 & 17 41 52.28 & $-$33 33 40.6 &    1.0 &    0.4 & 1.1~~~ & ~~~Yes & 10.484$\pm$0.029 & ~9.528$\pm$0.030 & ~9.018$\pm$0.029 \\
IRAS\,17499$-$3520 & 17 53 20.42 & $-$35 21 10.3 &    8.8 &    0.7 & 8.8~~~ & ~~~Yes & 10.663$\pm$0.024 & 10.005$\pm$0.027 & ~9.595$\pm$0.019 \\
IRAS\,17516$-$2525 & 17 54 43.35 & $-$25 26 28.0 & $-$2.0 & $-$1.0 & 2.3~~~ & ~~~Yes & ~8.695$\pm$0.020 & ~6.850$\pm$0.036 & ~5.082$\pm$0.016 \\
IRAS\,17540$-$2753 & 17 57 11.41 & $-$27 54 19.0 &$-$35.7 & $-$3.0 &35.8~~~ & ~~~No  & 12.326$\pm$0.028 & 10.903$\pm$0.029 & ~9.996$\pm$0.021 \\
IRAS\,17548$-$2753 & 17 57 58.27 & $-$27 53 20.2 &    7.6 & $-$0.2 & 7.6~~~ & ~~~No  & 15.3             & 12.78$\pm$0.04   & 11.465$\pm$0.032 \\
IRAS\,17550$-$2800 & 17 58 09.57 & $-$28 00 25.3 &$-$13.6 &    0.7 &13.7~~~ & ~~~Yes & 12.5             & 11.04$\pm$0.06   & ~8.9             \\
IRAS\,17580$-$3111 & 18 01 20.40 & $-$31 11 20.3 &    9.0 &    2.7 & 9.4~~~ & ~~~No  & 12.402$\pm$0.024 & 10.551$\pm$0.022 & ~9.280$\pm$0.019 \\
IRAS\,17582$-$2619 & 18 01 21.57 & $-$26 19 37.3 &    5.0 & $-$0.3 & 5.0~~~ & ~~~No  & 15.3             & 12.527$\pm$0.030 & 10.840$\pm$0.025 \\
IRAS\,17596$-$3952 & 18 03 06.77 & $-$39 51 53.8 &    6.6 &    1.2 & 6.7~~~ & ~~~Yes & ~9.624$\pm$0.026 & ~8.845$\pm$0.026 & ~8.410$\pm$0.022 \\
IRAS\,18015$-$1352 & 18 04 22.66 & $-$13 51 50.1 &    6.7 & $-$1.1 & 6.8~~~ & ~~~No  & 16.2             & 15.0             & 13.289$\pm$0.035 \\
IRAS\,18049$-$2118 & 18 07 55.12 & $-$21 18 08.3 &    4.5 &    0.7 & 4.5~~~ & ~~~No  & 16.6             & 13.9             & 11.312$\pm$0.035 \\
IRAS\,18071$-$1727 & 18 10 06.07 & $-$17 26 34.5 &    9.6 &   21.5 &23.5~~~ & ~~~No  & 15.9             & 14.1             & 12.71$\pm$0.06   \\
IRAS\,18105$-$1935 & 18 13 29.92 & $-$19 35 02.9 &$-$32.2 &    0.1 &32.2~~~ & ~~~No  & 11.3             & 10.81$\pm$0.04   & ~9.2             \\
IRAS\,18113$-$2503 & 18 14 27.26 & $-$25 03 00.4 &   14.4 & $-$5.4 &15.4~~~ & ~~~Yes & 11.630$\pm$0.033 & 10.698$\pm$0.034 & 10.323$\pm$0.030 \\
IRAS\,18183$-$2538 & 18 21 24.49 & $-$25 36 35.0 & $-$2.8 &    0.0 & 2.8~~~ & ~~~Yes & 14.59$\pm$0.04   & 13.043$\pm$0.026 & 11.344$\pm$0.023 \\
IRAS\,18199$-$1442 & 18 22 50.92 & $-$14 40 48.3 &    1.7 &    0.7 & 1.9~~~ & ~~~No  & 16.9             & 15.8             & 11.146$\pm$0.023 \\
IRAS\,18229$-$1127 & 18 25 41.49 & $-$11 26 14.5 &$-$51.6 &$-$18.5 &54.8~~~ & ~~~Yes & 12.416$\pm$0.029 & 11.39$\pm$0.04   & 10.696$\pm$0.029 \\
IRAS\,18236$-$0447 & 18 26 19.56 & $-$04 45 44.3 &$-$11.1 & $-$2.3 &11.3~~~ & ~~~No  & 16.5             & 15.30$\pm$0.11   & 13.10$\pm$0.04   \\
IRAS\,18355$-$0712 & 18 38 15.46 & $-$07 09 54.3 &    0.9 & $-$2.3 & 2.5~~~ & ~~~No  & 15.85$\pm$0.11   & 13.12$\pm$0.04   & 11.528$\pm$0.035 \\
IRAS\,18361$-$1203 & 18 38 59.12 & $-$12 00 42.0 &    4.7 &    2.0 & 5.1~~~ & ~~~Yes & 11.727$\pm$0.026 & 10.627$\pm$0.032 & 10.027$\pm$0.024 \\
IRAS\,18434$-$0042 & 18 46 04.70 & $-$00 38 54.4 &    4.5 &    0.6 & 4.5~~~ & ~~~No  & 16.13$\pm$0.10   & 13.806$\pm$0.028 & 12.573$\pm$0.022 \\
IRAS\,18454$+$0001 & 18 48 01.18 & $+$00 04 48.5 & $-$4.8 &    1.5 & 5.0~~~ & ~~~Yes & 14.061$\pm$0.026 & 12.876$\pm$0.024 & 12.112$\pm$0.025 \\
IRAS\,18470$+$0015 & 18 49 38.91 & $+$00 18 54.9 & $-$2.9 &    2.9 & 4.1~~~ & ~~~No  & 14.07$\pm$0.13   & 12.454$\pm$0.021 & 10.765$\pm$0.020 \\
IRAS\,18485$+$0642 & 18 50 59.60 & $+$06 45 59.0 &   10.4 &    4.0 &11.2~~~ & ~~~Yes & 14.020$\pm$0.030 & 12.949$\pm$0.031 & 12.325$\pm$0.029 \\
IRAS\,18524$+$0544 & 18 54 54.14 & $+$05 48 11.2 &    0.0 & $-$0.1 & 0.1~~~ & ~~~Yes & 14.58$\pm$0.04   & 13.70$\pm$0.04   & 12.412$\pm$0.033 \\
IRAS\,18576$+$0341 & 19 00 10.89 & $+$03 45 47.1 & $-$0.2 &    0.1 & 0.2~~~ & ~~~No  & 12.164$\pm$0.027 & ~8.918$\pm$0.028 & ~7.007$\pm$0.020 \\
IRAS\,19006$+$1022 & 19 03 00.03 & $+$10 26 36.8 &    1.9 &    1.8 & 2.6~~~ & ~~~No  & 16.34$\pm$0.13   & 13.55$\pm$0.04   & 11.440$\pm$0.023 \\
IRAS\,19075$+$0432 & 19 09 59.92 & $+$04 37 08.5 & $-$1.2 &    2.5 & 2.8~~~ & ~~~Yes & 10.973$\pm$0.021 & 10.061$\pm$0.024 & ~8.745$\pm$0.025 \\
IRAS\,19079$-$0315 & 19 10 32.66 & $-$03 10 13.0 &    3.9 &    3.0 & 4.9~~~ & ~~~Yes & 11.979$\pm$0.023 & 11.205$\pm$0.025 & 10.588$\pm$0.023 \\
IRAS\,19083$+$0119 & 19 10 54.53 & $+$01 24 45.0 &    1.9 &    3.0 & 3.6~~~ & ~~~No  & 16.9             & 15.5             & 14.34$\pm$0.08   \\
IRAS\,19094$+$1627 & 19 11 44.68 & $+$16 32 55.4 &    2.6 &    1.4 & 2.9~~~ & ~~~Yes & 12.829$\pm$0.033 & 11.8             & 11.1             \\
IRAS\,19134$+$2131 & 19 15 35.20 & $+$21 36 34.0 & $-$2.8 &    1.0 & 3.0~~~ & ~~~No  & 16.54$\pm$0.13   & 14.93$\pm$0.07   & 13.46$\pm$0.04   \\
IRAS\,19176$+$1251 & 19 19 55.77 & $+$12 57 38.2 & $-$0.4 &    4.2 & 4.2~~~ & ~~~No  & 16.04$\pm$0.09   & 14.52$\pm$0.05   & 13.020$\pm$0.026 \\
IRAS\,19181$+$1806 & 19 20 24.86 & $+$18 11 41.4 & $-$4.8 &    0.4 & 4.9~~~ & ~~~Yes & 13.472$\pm$0.025 & 12.086$\pm$0.018 & 10.984$\pm$0.020 \\
IRAS\,19193$+$1804 & 19 21 31.63 & $+$18 10 09.5 &    1.9 &    0.5 & 1.9~~~ & ~~~No  & 18.0             & 15.56$\pm$0.13   & 14.29$\pm$0.07   \\
IRAS\,19319$+$2214 & 19 34 03.58 & $+$22 21 15.9 &    0.0 & $-$0.1 & 0.1~~~ & ~~~No  & 13.668$\pm$0.024 & 12.003$\pm$0.022 & 10.747$\pm$0.018 \\
IRAS\,19374$+$2359 & 19 39 35.54 & $+$24 06 27.0 & $-$0.8 & $-$1.0 & 1.3~~~ & ~~~Yes & 12.038$\pm$0.025 & 10.866$\pm$0.023 & ~9.735$\pm$0.017 \\
IRAS\,20174$+$3222 & 20 19 27.81 & $+$32 32 15.2 & $-$2.4 &    2.2 & 3.3~~~ & ~~~Yes & 10.992$\pm$0.023 & 10.118$\pm$0.032 & ~9.620$\pm$0.021 \\
IRAS\,20214$+$3749 & 20 23 18.94 & $+$37 58 51.6 & $-$3.1 & $-$0.4 & 3.1~~~ & ~~~No  & 16.96$\pm$0.20   & 13.865$\pm$0.035 & 11.046$\pm$0.018 \\
IRAS\,20244$+$3509 & 20 26 25.16 & $+$35 19 13.4 & $-$2.9 & $-$0.6 & 3.0~~~ & ~~~Yes & ~9.522$\pm$0.020 & ~8.188$\pm$0.029 & ~7.394$\pm$0.017 \\
IRAS\,20406$+$2953 & 20 42 45.95 & $+$30 04 06.4 & $-$0.2 & $-$0.1 & 0.2~~~ & ~~~Yes & 10.966$\pm$0.021 & ~9.065$\pm$0.023 & ~7.920$\pm$0.021 \\
IRAS\,20461$+$3853 & 20 48 04.43 & $+$39 04 59.5 & $-$2.0 & $-$0.5 & 2.0~~~ & ~~~Yes & 11.451$\pm$0.021 & 10.294$\pm$0.019 & ~9.664$\pm$0.015 \\
IRAS\,21525$+$5643 & 21 54 14.58 & $+$56 57 27.0 & $-$3.4 &    4.0 & 5.3~~~ & ~~~No  & 16.5             & 16.00$\pm$0.16   & 13.50$\pm$0.07   \\

\hline
\end{tabular}}
\end{table*}


\begin{table*}
\footnotesize{
\caption{\footnotesize{
\emph{IRAS} Post-AGB Star and PN Candidates with no Counterparts in the 
2MASS PSC.  }}

\label{table:3}      
\centering          
\begin{tabular}{lll}     
\hline\hline       

\multicolumn{1}{c}{Object} & 
\multicolumn{1}{c}{Optical} & 
\multicolumn{1}{l}{Comments} \\

\hline 

IRAS\,04137$+$7016 & ~~~Yes & Two objects unresolved by 2MASS. \\
IRAS\,05573$+$3156 & ~~~No  & Diffuse emission. \\
IRAS\,06499$+$0145 & ~~~Yes & Diffuse emission. \\
IRAS\,11381$-$6401 & ~~~Yes & Two objects unresolved by 2MASS. \\
IRAS\,11544$-$6408 & ~~~Yes & Source very faint. 2MASS magnitudes not available. \\
IRAS\,12360$-$5740 & ~~~Yes & 2MASS magnitudes contaminated by a nearby star. \\
IRAS\,12405$-$6219 & ~~~No  & Diffuse emission or multiple objects unresolved by 2MASS. \\
IRAS\,13293$-$6000 & ~~~No  & Two objects unresolved by 2MASS. \\
IRAS\,13427$-$6531 & ~~~Yes & Multiple objects unresolved by 2MASS. \\
IRAS\,13529$-$5934 & ~~~Yes & Two objects unresolved by 2MASS. \\
IRAS\,14249$-$5310 & ~~~No  & Two objects unresolved by 2MASS. \\
IRAS\,15103$-$5754 & ~~~No  & Diffuse emission. \\
IRAS\,15229$-$5433 & ~~~Yes & Diffuse emission or multiple objects unresolved by 2MASS. \\
IRAS\,15452$-$5459 & ~~~No  & Diffuse emission. \\ 
IRAS\,15534$-$5422 & ~~~Yes & Diffuse emission. \\
IRAS\,16209$-$4714 & ~~~Yes & 2MASS magnitudes contaminated by a nearby star. \\
IRAS\,16296$-$4507 & ~~~Yes & Diffuse emission or two objects unresolved by 2MASS. \\
IRAS\,16333$-$4807 & ~~~No  & Diffuse emission. \\
IRAS\,16507$-$4810 & ~~~Yes & Diffuse emission or two objects unresolved by 2MASS. \\
IRAS\,17009$-$4154 & ~~~No  & Diffuse emission. \\
IRAS\,17052$-$3245 & ~~~No  & Source very faint. 2MASS magnitudes not available. \\
IRAS\,17150$-$3224 & ~~~Yes & Diffuse emission, the Cotton Candy Nebula. \\
IRAS\,17234$-$4008 & ~~~No  & Two objects unresolved by 2MASS. \\
IRAS\,17269$-$2235 & ~~~Yes & Two objects unresolved by 2MASS. \\
IRAS\,17301$-$2538 & ~~~Yes & Two objects unresolved by 2MASS. \\
IRAS\,17361$-$4159 & ~~~No  & Two objects unresolved by 2MASS. \\
IRAS\,17376$-$3448 & ~~~No  & Diffuse emission or multiple objects unresolved by 2MASS. \\
IRAS\,17418$-$3335 & ~~~No  & Two objects unresolved by 2MASS. \\
IRAS\,17479$-$3032 & ~~~No  & Two objects unresolved by 2MASS. \\
IRAS\,17487$-$1922 & ~~~Yes & Two objects unresolved by 2MASS. \\
IRAS\,17506$-$2955 & ~~~No  & Two objects unresolved by 2MASS. \\
IRAS\,17560$-$2027 & ~~~Yes & Two objects unresolved by 2MASS. \\
IRAS\,18087$-$1440 & ~~~No  & Two objects unresolved by 2MASS. \\
IRAS\,19015$+$1256 & ~~~No  & Two objects unresolved by 2MASS. \\
IRAS\,19071$+$0857 & ~~~No  & Diffuse emission or multiple objects unresolved by 2MASS. \\
IRAS\,19178$+$1206 & ~~~No  & Two objects unresolved by 2MASS. \\
IRAS\,19208$+$1541 & ~~~No  & Two objects unresolved by 2MASS. \\
IRAS\,20035$+$3242 & ~~~No  & Two objects unresolved by 2MASS. \\
IRAS\,21554$+$6204 & ~~~No  & Two objects unresolved by 2MASS. \\

\hline
\end{tabular}}
\end{table*}


\begin{table*}
\footnotesize{
\caption{\footnotesize{
\emph{IRAS} Post-AGB Star and PN Candidates Without 2MASS Near-IR 
Counterparts.  }}

\label{table:4}
\centering          
\begin{tabular}{l|l|l|l|l|l} 
\hline\hline       



IRAS\,08351$-$4634 & IRAS\,15531$-$5704 & IRAS\,17158$-$4049 & IRAS\,17543$-$3102 & IRAS\,18135$-$1456 & IRAS\,19011$+$1049 \\
IRAS\,09370$-$4826 & IRAS\,16245$-$3859 & IRAS\,17168$-$3736 & IRAS\,17550$-$2120 & IRAS\,18246$-$1032 & IRAS\,19013$+$0629 \\
IRAS\,10194$-$5625 & IRAS\,16518$-$3425 & IRAS\,17382$-$2531 & IRAS\,17552$-$2030 & IRAS\,18385$+$1350 & IRAS\,19190$+$1102 \\
IRAS\,11444$-$6150 & IRAS\,16558$-$3417 & IRAS\,17385$-$2413 & IRAS\,18011$-$1847 & IRAS\,18514$+$0019 & IRAS\,19315$+$2235 \\
IRAS\,11549$-$6225 & IRAS\,17067$-$3759 & IRAS\,17393$-$2727 & IRAS\,18016$-$2743 & IRAS\,18529$+$0210 & IRAS\,19454$+$2920 \\
IRAS\,13404$-$6059 & IRAS\,17088$-$4221 & IRAS\,17404$-$2713 & IRAS\,18039$-$1903 & IRAS\,18580$+$0818 & IRAS\,20042$+$3259 \\
IRAS\,13500$-$6106 & IRAS\,17097$-$3624 & IRAS\,17443$-$2949 & IRAS\,18051$-$2415 & IRAS\,18596$+$0315 & IRAS\,21537$+$6435 \\
IRAS\,15284$-$6026 & IRAS\,17153$-$3814 & IRAS\,17482$-$2501 & IRAS\,18083$-$2155 &                    &                    \\

\hline
\end{tabular}}
\end{table*}


\begin{table*}
\scriptsize{
\caption{\footnotesize{
\emph{Spitzer} and \emph{MSX} Fluxes of \emph{IRAS} Post-AGB Star and PN Candidates with Counterparts 
in the 2MASS PSC.}}           
\label{table:5}      
\centering       
\begin{tabular}{rccccccccc}     
\hline\hline 
\multicolumn{1}{c}{} & 
\multicolumn{4}{c}{Spitzer} & 
\multicolumn{1}{c}{} & 
\multicolumn{4}{c}{MSX} \\ 
\cline{2-5}
\cline{7-10} 
\\
\multicolumn{1}{r}{} & 
\multicolumn{1}{c}{3.6 $\mu$m} & 
\multicolumn{1}{c}{4.5 $\mu$m} & 
\multicolumn{1}{c}{5.8 $\mu$m} & 
\multicolumn{1}{c}{8 $\mu$m} & 
\multicolumn{1}{c}{} & 
\multicolumn{1}{c}{8.28 $\mu$m} & 
\multicolumn{1}{c}{12.13 $\mu$m} & 
\multicolumn{1}{c}{14.65 $\mu$m} & 
\multicolumn{1}{c}{21.3 $\mu$m} \\ 
\multicolumn{1}{c}{} & 
\multicolumn{1}{c}{[{\rm Jy}]} & 
\multicolumn{1}{c}{[{\rm Jy}]} & 
\multicolumn{1}{c}{[{\rm Jy}]} & 
\multicolumn{1}{c}{[{\rm Jy}]} & 
\multicolumn{1}{c}{} & 
\multicolumn{1}{c}{[{\rm Jy}]} & 
\multicolumn{1}{c}{[{\rm Jy}]} & 
\multicolumn{1}{c}{[{\rm Jy}]} & 
\multicolumn{1}{c}{[{\rm Jy}]} \\

\hline 
 
IRAS\,00509$+$6623 &   $\cdots$~~~~~ &   $\cdots$~~~~~ &   $\cdots$~~~~~ &   $\cdots$~~~~~ & & ~1.18$\pm$0.05 & 2.75$\pm$0.19 & 2.45$\pm$0.18 & 3.50$\pm$0.27 \\
IRAS\,08242$-$3828 &   $\cdots$~~~~~ &   $\cdots$~~~~~ &   $\cdots$~~~~~ &   $\cdots$~~~~~ & & ~8.9$\pm$0.4   & 11.3$\pm$0.6  & 12.9$\pm$0.8  & 20.7$\pm$1.2  \\
IRAS\,09055$-$4629 &   $\cdots$~~~~~ &   $\cdots$~~~~~ &   $\cdots$~~~~~ &   $\cdots$~~~~~ & & ~0.26$\pm$0.01 & 0.78$\pm$0.05 & 0.86$\pm$0.06 & 1.98$\pm$0.13 \\
IRAS\,09102$-$5101 &   $\cdots$~~~~~ &   $\cdots$~~~~~ &   $\cdots$~~~~~ &   $\cdots$~~~~~ & & ~0.14$\pm$0.01 & $\cdots$~~~~~ & 1.06$\pm$0.08 & 1.63$\pm$0.13 \\
IRAS\,09119$-$5150 &   $\cdots$~~~~~ &   $\cdots$~~~~~ &   $\cdots$~~~~~ &   $\cdots$~~~~~ & & ~0.30$\pm$0.02 & 1.05$\pm$0.09 & 1.26$\pm$0.09 & 2.14$\pm$0.16 \\
IRAS\,09500$-$5236 &   $\cdots$~~~~~ &   $\cdots$~~~~~ &   $\cdots$~~~~~ &   $\cdots$~~~~~ & & ~0.23$\pm$0.01 & $\cdots$~~~~~ & 0.88$\pm$0.07 & $\cdots$~~~~~ \\
IRAS\,09378$-$5117 &   $\cdots$~~~~~ &   $\cdots$~~~~~ &   $\cdots$~~~~~ &   $\cdots$~~~~~ & & ~0.76$\pm$0.03 & 2.17$\pm$0.13 & 3.89$\pm$0.24 & 4.92$\pm$0.31 \\
IRAS\,11339$-$6004 &   $\cdots$~~~~~ &   $\cdots$~~~~~ &   $\cdots$~~~~~ &   $\cdots$~~~~~ & & ~0.67$\pm$0.03 & 2.37$\pm$0.14 & 3.28$\pm$0.21 & 5.65$\pm$0.35 \\
IRAS\,11488$-$6432 &   $\cdots$~~~~~ &   $\cdots$~~~~~ &   $\cdots$~~~~~ &   $\cdots$~~~~~ & & ~1.03$\pm$0.04 & 1.49$\pm$0.10 & 1.72$\pm$0.11 & 2.02$\pm$0.15 \\
IRAS\,11544$-$6408 &   $\cdots$~~~~~ &   $\cdots$~~~~~ &   $\cdots$~~~~~ &   $\cdots$~~~~~ & & 10.34$\pm$0.42 & 24.6$\pm$1.2  & 34.1$\pm$2.1  & 34.0$\pm$2.0  \\
IRAS\,12262$-$6417 &   $\cdots$~~~~~ &   $\cdots$~~~~~ &   $\cdots$~~~~~ &   $\cdots$~~~~~ & & ~1.86$\pm$0.08 & 2.55$\pm$0.15 & 2.64$\pm$0.17 & 5.01$\pm$0.32 \\
IRAS\,12309$-$5928 &   $\cdots$~~~~~ &   $\cdots$~~~~~ &   $\cdots$~~~~~ &   $\cdots$~~~~~ & & ~1.31$\pm$0.05 & 2.00$\pm$0.12 & 2.34$\pm$0.15 & 3.90$\pm$0.25 \\
IRAS\,13356$-$6249 &   $\cdots$~~~~~ &   $\cdots$~~~~~ & 0.747$\pm$0.022 & 0.882$\pm$0.030 & & ~1.49$\pm$0.06 & 7.6$\pm$0.4   & 22.1$\pm$1.4  &  103$\pm$6    \\
IRAS\,13398$-$5951 &   $\cdots$~~~~~ &   $\cdots$~~~~~ &   $\cdots$~~~~~ &   $\cdots$~~~~~ & & ~1.57$\pm$0.06 & 2.70$\pm$0.15 & 3.37$\pm$0.21 & 4.81$\pm$0.30 \\
IRAS\,13421$-$6125 & 0.273$\pm$0.031 &   $\cdots$~~~~~ &   $\cdots$~~~~~ &   $\cdots$~~~~~ & & ~7.14$\pm$0.29 & 14.6$\pm$0.7  & 22.4$\pm$1.4  & 22.0$\pm$1.3  \\
IRAS\,13483$-$5905 &   $\cdots$~~~~~ &   $\cdots$~~~~~ &   $\cdots$~~~~~ &   $\cdots$~~~~~ & & ~1.07$\pm$0.04 & 4.45$\pm$0.23 & 10.0$\pm$0.6  & 10.8$\pm$0.7  \\
IRAS\,14104$-$5819 &   $\cdots$~~~~~ &   $\cdots$~~~~~ &   $\cdots$~~~~~ &   $\cdots$~~~~~ & & ~3.64$\pm$0.15 & 4.59$\pm$0.24 &  7.8$\pm$0.5  &  9.2$\pm$0.6  \\
IRAS\,15038$-$5533 &   $\cdots$~~~~~ &   $\cdots$~~~~~ &   $\cdots$~~~~~ &   $\cdots$~~~~~ & & ~0.63$\pm$0.03 & 1.25$\pm$0.09 & 1.61$\pm$0.11 & 2.15$\pm$0.16 \\
IRAS\,15144$-$5812 &  0.35$\pm$0.07  &  3.35$\pm$0.21  &  4.41$\pm$0.11  &   $\cdots$~~~~~ & & ~5.13$\pm$0.21 & 6.53$\pm$0.33 & 8.8$\pm$0.5   & 26.6$\pm$1.6  \\
IRAS\,15408$-$5657 &   $\cdots$~~~~~ &   $\cdots$~~~~~ &   $\cdots$~~~~~ &   $\cdots$~~~~~ & &  65.9$\pm$2.7  &  102$\pm$5    & 112$\pm$7     &  123$\pm$7    \\
IRAS\,15408$-$5413 &   $\cdots$~~~~~ &   $\cdots$~~~~~ &   $\cdots$~~~~~ &   $\cdots$~~~~~ & &   135$\pm$6~   &  231$\pm$12   & 257$\pm$16    &  298$\pm$18   \\
IRAS\,16228$-$5014 &   $\cdots$~~~~~ &   $\cdots$~~~~~ & 0.690$\pm$0.020 & 0.463$\pm$0.011 & & ~0.22$\pm$0.01 & 0.76$\pm$0.07 & 1.35$\pm$0.09 & 11.4$\pm$0.7  \\
IRAS\,16517$-$3626 &   $\cdots$~~~~~ &   $\cdots$~~~~~ &   $\cdots$~~~~~ &   $\cdots$~~~~~ & & ~0.75$\pm$0.03 & 1.65$\pm$0.13 & 2.39$\pm$0.16 & 5.45$\pm$0.35 \\
IRAS\,16518$-$3425 &   $\cdots$~~~~~ &   $\cdots$~~~~~ &   $\cdots$~~~~~ &   $\cdots$~~~~~ & & ~0.35$\pm$0.02 & $\cdots$~~~~~ & 1.95$\pm$0.14 &  6.7$\pm$0.4  \\
IRAS\,16567$-$3838 &   $\cdots$~~~~~ &   $\cdots$~~~~~ &   $\cdots$~~~~~ &   $\cdots$~~~~~ & & ~5.88$\pm$0.24 & 8.5$\pm$0.4   & 11.1$\pm$0.7  & 13.4$\pm$0.8  \\
IRAS\,16584$-$3710 &   $\cdots$~~~~~ &   $\cdots$~~~~~ &   $\cdots$~~~~~ &   $\cdots$~~~~~ & & ~0.43$\pm$0.02 & 0.95$\pm$0.10 & 1.96$\pm$0.13 & 4.00$\pm$0.26 \\
IRAS\,17010$-$3810 &   $\cdots$~~~~~ &   $\cdots$~~~~~ &   $\cdots$~~~~~ &   $\cdots$~~~~~ & & ~0.55$\pm$0.02 & 2.21$\pm$0.13 & 4.08$\pm$0.25 &  8.2$\pm$0.5  \\
IRAS\,17021$-$3109 &   $\cdots$~~~~~ &   $\cdots$~~~~~ &   $\cdots$~~~~~ &   $\cdots$~~~~~ & & ~1.06$\pm$0.05 & 2.99$\pm$0.21 &  6.3$\pm$0.4  &  9.1$\pm$0.6  \\
IRAS\,17021$-$3054 &   $\cdots$~~~~~ &   $\cdots$~~~~~ &   $\cdots$~~~~~ &   $\cdots$~~~~~ & & ~0.42$\pm$0.02 & $\cdots$~~~~~ & 3.32$\pm$0.23 & 4.51$\pm$0.32 \\
IRAS\,17149$-$3053 &   $\cdots$~~~~~ &   $\cdots$~~~~~ &   $\cdots$~~~~~ &   $\cdots$~~~~~ & & ~0.41$\pm$0.02 & 2.02$\pm$0.13 & 3.93$\pm$0.25 &  8.2$\pm$0.5  \\
IRAS\,17175$-$2819 &   $\cdots$~~~~~ &   $\cdots$~~~~~ &   $\cdots$~~~~~ &   $\cdots$~~~~~ & &  12.7$\pm$0.5  &   61$\pm$3    &  128$\pm$8    &  247$\pm$15   \\
IRAS\,17233$-$2602 &   $\cdots$~~~~~ &   $\cdots$~~~~~ &   $\cdots$~~~~~ &   $\cdots$~~~~~ & & ~1.94$\pm$0.08 & 4.32$\pm$0.23 &  7.4$\pm$0.5  &  7.9$\pm$0.5  \\
IRAS\,17348$-$2906 &   $\cdots$~~~~~ &   $\cdots$~~~~~ &  2.54$\pm$0.07  &   $\cdots$~~~~~ & & ~1.92$\pm$0.08 & 3.30$\pm$0.18 & 3.92$\pm$0.24 &  7.7$\pm$0.5  \\
IRAS\,17359$-$2902 &0.0520$\pm$0.0020&0.0522$\pm$0.0017&0.0610$\pm$0.0024& 0.261$\pm$0.007 & & ~0.44$\pm$0.02 & 1.70$\pm$0.11 & 3.27$\pm$0.21 & 10.0$\pm$0.6  \\
IRAS\,17360$-$2142 &   $\cdots$~~~~~ &   $\cdots$~~~~~ &   $\cdots$~~~~~ &   $\cdots$~~~~~ & & ~0.26$\pm$0.01 &  1.3$\pm$0.6  & 2.31$\pm$0.15 &  8.9$\pm$0.5  \\
IRAS\,17385$-$3332 &   $\cdots$~~~~~ &   $\cdots$~~~~~ &   $\cdots$~~~~~ &   $\cdots$~~~~~ & & ~0.19$\pm$0.01 & 0.73$\pm$0.08 & 1.78$\pm$0.12 &  7.7$\pm$0.5  \\
IRAS\,17499$-$3520 &   $\cdots$~~~~~ &   $\cdots$~~~~~ &   $\cdots$~~~~~ &   $\cdots$~~~~~ & & ~0.76$\pm$0.03 & 2.40$\pm$0.15 & 3.65$\pm$0.23 & 10.3$\pm$0.6  \\
IRAS\,17516$-$2525 &   $\cdots$~~~~~ &   $\cdots$~~~~~ &   $\cdots$~~~~~ &   $\cdots$~~~~~ & &  40.5$\pm$1.7  & 43.9$\pm$2.2  & 44.7$\pm$2.7  &   89$\pm$5    \\
IRAS\,17540$-$2753 &0.0590$\pm$0.0020&0.0483$\pm$0.0027&0.0580$\pm$0.0018& 0.292$\pm$0.010 & & ~0.38$\pm$0.02 & 1.41$\pm$0.10 & 3.74$\pm$0.24 & 10.0$\pm$0.6  \\
IRAS\,17548$-$2753 &0.0367$\pm$0.0015&0.0462$\pm$0.0019&0.0480$\pm$0.0014& 0.136$\pm$0.004 & & ~0.23$\pm$0.01 & 1.00$\pm$0.09 & 3.60$\pm$0.23 & 11.4$\pm$0.7  \\
IRAS\,17550$-$2800 &   $\cdots$~~~~~ &   $\cdots$~~~~~ &  1.90$\pm$0.06  &   $\cdots$~~~~~ & & ~2.29$\pm$0.09 & 2.17$\pm$0.13 & 3.00$\pm$0.19 & 4.28$\pm$0.28 \\
IRAS\,17580$-$3111 &   $\cdots$~~~~~ &   $\cdots$~~~~~ &   $\cdots$~~~~~ &   $\cdots$~~~~~ & & ~1.24$\pm$0.05 & 3.62$\pm$0.18 &  6.9$\pm$0.4  & 13.2$\pm$0.8  \\
IRAS\,17582$-$2619 &   $\cdots$~~~~~ &   $\cdots$~~~~~ &   $\cdots$~~~~~ &   $\cdots$~~~~~ & & ~0.38$\pm$0.02 & 1.65$\pm$0.11 & 3.45$\pm$0.22 &  7.0$\pm$0.4  \\
IRAS\,18015$-$1352 &   $\cdots$~~~~~ &   $\cdots$~~~~~ &   $\cdots$~~~~~ &   $\cdots$~~~~~ & & ~1.46$\pm$0.06 & 2.39$\pm$0.14 & 2.48$\pm$0.16 & 3.42$\pm$0.22 \\
IRAS\,18049$-$2118 &   $\cdots$~~~~~ &  3.90$\pm$0.34  &  11.4$\pm$0.8   &   $\cdots$~~~~~ & & ~8.31$\pm$0.34 & 13.4$\pm$0.7  & 17.6$\pm$1.1  & 18.3$\pm$1.1  \\
IRAS\,18071$-$1727 & 0.147$\pm$0.006 &   $\cdots$~~~~~ &   $\cdots$~~~~~ &   $\cdots$~~~~~ & & ~8.22$\pm$0.34 & 28.7$\pm$1.4  & 51.5$\pm$3.1  & 57.5$\pm$3.5  \\
IRAS\,18105$-$1935 & 0.370$\pm$0.021 &   $\cdots$~~~~~ &  2.22$\pm$0.05  &  4.32$\pm$0.11  & & ~3.78$\pm$0.15 &  8.3$\pm$0.4  & 12.0$\pm$0.7  & 15.0$\pm$0.9  \\
IRAS\,18113$-$2503 &   $\cdots$~~~~~ &   $\cdots$~~~~~ &   $\cdots$~~~~~ &   $\cdots$~~~~~ & & ~0.69$\pm$0.03 & 2.53$\pm$0.14 & 5.48$\pm$0.33 &  9.6$\pm$0.6  \\
IRAS\,18199$-$1442 & 0.542$\pm$0.04  &  2.83$\pm$0.15  &   $\cdots$~~~~~ &   $\cdots$~~~~~ & &  15.7$\pm$0.6  & 22.3$\pm$1.1  & 29.8$\pm$1.8  & 28.4$\pm$1.7  \\
IRAS\,18229$-$1127 &0.0368$\pm$0.0029&0.0434$\pm$0.0024& 0.161$\pm$0.007 & 0.515$\pm$0.04  & & ~0.50$\pm$0.02 & 1.18$\pm$0.09 & 1.92$\pm$0.12 & 17.4$\pm$1.1  \\
IRAS\,18236$-$0447 &   $\cdots$~~~~~ &   $\cdots$~~~~~ &   $\cdots$~~~~~ &   $\cdots$~~~~~ & & ~0.67$\pm$0.03 & 1.40$\pm$0.10 & 2.70$\pm$0.17 & 2.73$\pm$0.19 \\
IRAS\,18355$-$0712 &0.0416$\pm$0.0019&0.0689$\pm$0.0019&0.1011$\pm$0.0021&0.227$\pm$0.004  & & ~0.27$\pm$0.01 & 1.55$\pm$0.08 & 3.86$\pm$0.24 &  9.4$\pm$0.6  \\
IRAS\,18361$-$1203 &   $\cdots$~~~~~ &   $\cdots$~~~~~ &   $\cdots$~~~~~ &   $\cdots$~~~~~ & & ~1.57$\pm$0.06 & 2.63$\pm$0.15 & 3.29$\pm$0.21 & 4.31$\pm$0.27 \\
IRAS\,18434$-$0042 &0.0078$\pm$0.0004&0.0113$\pm$0.0007&0.0266$\pm$0.0010&0.1249$\pm$0.0031& & ~0.14$\pm$0.01 & 0.62$\pm$0.07 & 1.48$\pm$0.10 & 2.81$\pm$0.18 \\
IRAS\,18454$+$0001 &0.0105$\pm$0.0004&0.0106$\pm$0.0003&0.0319$\pm$0.0009& 0.099$\pm$0.004 & & ~0.12$\pm$0.01 & 0.41$\pm$0.03 & 1.24$\pm$0.08 &  9.5$\pm$0.6  \\
IRAS\,18470$+$0015 & 0.216$\pm$0.009 & 0.447$\pm$0.018 & 0.807$\pm$0.022 &   $\cdots$~~~~~ & & ~2.15$\pm$0.09 & 4.64$\pm$0.23 &  6.1$\pm$0.4  &  8.8$\pm$0.5  \\
IRAS\,18485$+$0642 &   $\cdots$~~~~~ &   $\cdots$~~~~~ &   $\cdots$~~~~~ &   $\cdots$~~~~~ & & ~0.99$\pm$0.04 & 3.42$\pm$0.18 &  7.8$\pm$0.5  & 14.9$\pm$0.9  \\
IRAS\,18524$+$0544 &   $\cdots$~~~~~ &   $\cdots$~~~~~ &   $\cdots$~~~~~ &   $\cdots$~~~~~ & & ~0.19$\pm$0.01 & $\cdots$~~~~~ & 1.44$\pm$0.10 & 3.43$\pm$0.22 \\
IRAS\,18576$+$0341 &   2.3$\pm$0.4   &   $\cdots$~~~~~ &   $\cdots$~~~~~ &   $\cdots$~~~~~ & &  10.6$\pm$0.4  & 45.5$\pm$2.3  &   89$\pm$5    &  277$\pm$17   \\
IRAS\,19006$+$1022 &   $\cdots$~~~~~ &   $\cdots$~~~~~ &   $\cdots$~~~~~ &   $\cdots$~~~~~ & & ~0.56$\pm$0.02 & $\cdots$~~~~~ & 1.47$\pm$0.10 & 2.93$\pm$0.20 \\
IRAS\,19075$+$0432 &   $\cdots$~~~~~ &   $\cdots$~~~~~ &   $\cdots$~~~~~ &   $\cdots$~~~~~ & & ~4.51$\pm$0.19 & 5.34$\pm$0.27 &  8.6$\pm$0.5  & 17.0$\pm$1.0  \\
IRAS\,19083$+$0119 &   $\cdots$~~~~~ &   $\cdots$~~~~~ &   $\cdots$~~~~~ &   $\cdots$~~~~~ & & ~0.54$\pm$0.02 & 2.18$\pm$0.13 & 5.08$\pm$0.31 & 11.1$\pm$0.7  \\
IRAS\,19094$+$1627 &   $\cdots$~~~~~ &   $\cdots$~~~~~ &   $\cdots$~~~~~ &   $\cdots$~~~~~ & & ~0.32$\pm$0.02 & 0.75$\pm$0.08 & 1.34$\pm$0.10 & 2.40$\pm$0.17 \\
IRAS\,19134$+$2131 &   $\cdots$~~~~~ &   $\cdots$~~~~~ &   $\cdots$~~~~~ &   $\cdots$~~~~~ & & ~1.22$\pm$0.05 & 5.06$\pm$0.27 &  8.9$\pm$0.5  & 10.1$\pm$0.6  \\
IRAS\,19176$+$1251 &0.0071$\pm$0.0003&0.0124$\pm$0.0005&0.0189$\pm$0.0006& 0.162$\pm$0.005 & & ~0.32$\pm$0.01 & 1.13$\pm$0.08 & 2.06$\pm$0.13 &  8.1$\pm$0.5  \\
IRAS\,19181$+$1806 &   $\cdots$~~~~~ &   $\cdots$~~~~~ &   $\cdots$~~~~~ &   $\cdots$~~~~~ & & ~0.22$\pm$0.01 & $\cdots$~~~~~ & 0.80$\pm$0.07 & 2.63$\pm$0.18 \\
IRAS\,19193$+$1804 &   $\cdots$~~~~~ &   $\cdots$~~~~~ &   $\cdots$~~~~~ &   $\cdots$~~~~~ & & ~0.11$\pm$0.01 & $\cdots$~~~~~ & 1.95$\pm$0.13 & 5.17$\pm$0.32 \\
IRAS\,19319$+$2214 &   $\cdots$~~~~~ &   $\cdots$~~~~~ &   $\cdots$~~~~~ &   $\cdots$~~~~~ & & ~2.23$\pm$0.09 & 3.52$\pm$0.19 & 4.82$\pm$0.30 & 5.27$\pm$0.33 \\
IRAS\,19374$+$2359 & 0.430$\pm$0.011 &   $\cdots$~~~~~ &   $\cdots$~~~~~ &   $\cdots$~~~~~ & &  13.8$\pm$0.6  &23.72$\pm$1.19 & 33.7$\pm$2.1  &   68$\pm$4    \\
IRAS\,20174$+$3222 &   $\cdots$~~~~~ &   $\cdots$~~~~~ &   $\cdots$~~~~~ &   $\cdots$~~~~~ & & ~0.87$\pm$0.04 & 2.81$\pm$0.16 & 3.32$\pm$0.21 & 11.5$\pm$0.7  \\
IRAS\,20214$+$3749 &   $\cdots$~~~~~ &   $\cdots$~~~~~ &   $\cdots$~~~~~ &   $\cdots$~~~~~ & & ~1.23$\pm$0.05 & 1.93$\pm$0.12 & 2.18$\pm$0.14 & 2.60$\pm$0.18 \\
IRAS\,20244$+$3509 &   $\cdots$~~~~~ &   $\cdots$~~~~~ &   $\cdots$~~~~~ &   $\cdots$~~~~~ & & ~0.76$\pm$0.03 & 1.08$\pm$0.09 & 1.53$\pm$0.11 & 3.03$\pm$0.20 \\
IRAS\,20461$+$3853 &   $\cdots$~~~~~ &   $\cdots$~~~~~ &   $\cdots$~~~~~ &   $\cdots$~~~~~ & & ~0.82$\pm$0.03 & 2.48$\pm$0.14 & 3.01$\pm$0.19 &  6.8$\pm$0.4  \\
IRAS\,21525$+$5643 &   $\cdots$~~~~~ &   $\cdots$~~~~~ &   $\cdots$~~~~~ &   $\cdots$~~~~~ & & ~2.28$\pm$0.09 & 5.25$\pm$0.27 &  6.8$\pm$0.4  &  9.9$\pm$0.6  \\

\hline                
\end{tabular}}
\end{table*}

\end{document}